\documentclass[letterpaper,twocolumn, 10pt,accepted=2024-03-03]{quantumarticle}
\pdfoutput=1
\usepackage[utf8]{inputenc}
\usepackage[english]{babel}
\usepackage[T1]{fontenc}
\usepackage{amsmath}
\usepackage{hyperref}
\usepackage{upgreek}
\usepackage{tikz}
\usepackage{lipsum}
\usepackage{amsthm}

\theoremstyle{definition}
\newtheorem{implication}{Implication}

\addto\extrasenglish{

}
\usepackage{array}
\newcolumntype{C}[1]{>{\centering\arraybackslash}m{#1}}

\usepackage{color,soul}
\usepackage{xcolor}
\usepackage{listings}
\lstdefinestyle{base}{
    emptylines=1,
    breaklines=true,
    basicstyle=\ttfamily\color{black},
    columns=fullflexible,
    keepspaces=true,
    moredelim=**[is][\color{lightgray}]{@}{@},
}

\begin{document}

\title{Compiling Quantum Circuits for Dynamically Field-Programmable Neutral Atoms Array Processors}

\author{Daniel Bochen Tan}
\affiliation{Computer Science Department, University of California, Los Angeles, CA 90095}
\thanks{an extended abstract of this work was presented at the 41st International Conference on Computer-Aided Design (ICCAD '22~\cite{iccad22-olsq-raa}).}
\email{bochentan@g.ucla.edu}
\orcid{0000-0002-9711-2441}
\author{Dolev Bluvstein}
\email{dbluvstein@g.harvard.edu}
\author{Mikhail D. Lukin}
\email{lukin@physics.harvard.edu}
\affiliation{Department of Physics, Harvard University, Cambridge, MA 02138}
\author{Jason Cong}
\email{cong@cs.ucla.edu}
\affiliation{Computer Science Department, University of California, Los Angeles, CA 90095}
\orcid{0000-0003-2887-6963}

\maketitle

\begin{abstract}
Dynamically field-programmable qubit arrays (DPQA) have recently emerged as a promising platform for quantum information processing.
In DPQA, atomic qubits are selectively loaded into arrays of optical traps that can be reconfigured during the computation itself.
Leveraging qubit transport and parallel, entangling quantum operations, different pairs of qubits, even those initially far away, can be entangled at different stages of the quantum program execution.
Such reconfigurability and non-local connectivity present new challenges for compilation, especially in the layout synthesis step which places and routes the qubits and schedules the gates.
In this paper, we consider a DPQA architecture that contains multiple arrays and supports 2D array movements, representing cutting-edge experimental platforms.
Within this architecture, we discretize the state space  and formulate layout synthesis as a satisfiability modulo theories problem, which can be solved by existing solvers optimally in terms of circuit depth.
For a set of benchmark circuits generated by random graphs with complex connectivities, our compiler OLSQ-DPQA reduces the number of two-qubit entangling gates on small problem instances by 1.7x compared to optimal compilation results on a fixed planar architecture.
To further improve scalability and practicality of the method, we introduce a greedy heuristic inspired by the iterative peeling approach in classical integrated circuit routing.
Using a hybrid approach that combined the greedy and optimal methods, we demonstrate that our DPQA-based compiled circuits feature reduced scaling overhead compared to a grid fixed architecture, resulting in 5.1X less two-qubit gates for 90 qubit quantum circuits.
These methods enable programmable, complex quantum circuits with neutral atom quantum computers, as well as informing both future compilers and future hardware choices.
\end{abstract}

\begin{figure*}
    \centering
        \includegraphics[width=\linewidth]{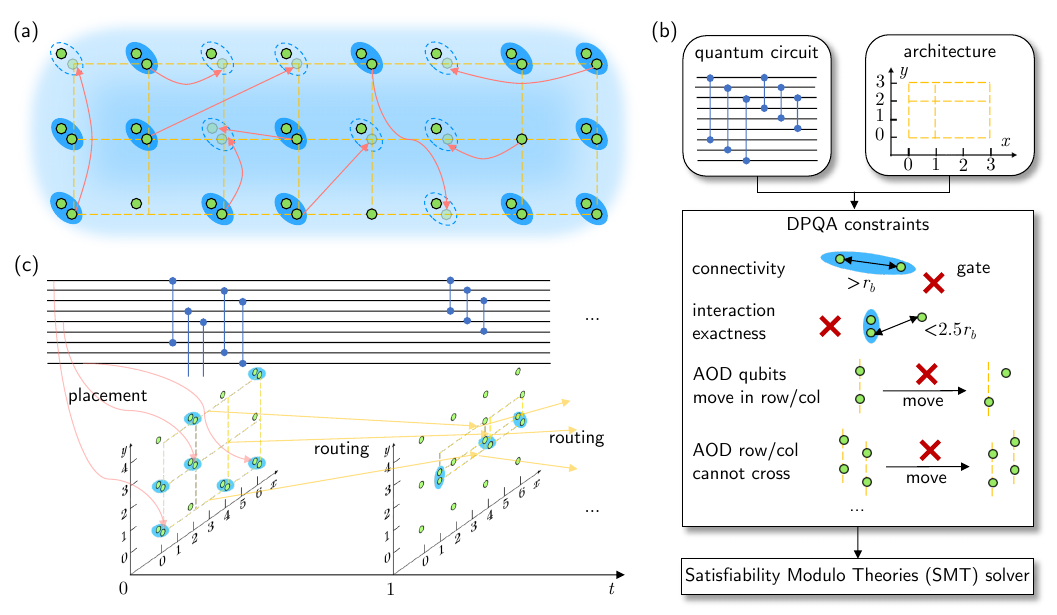}
        \caption{Compiling quantum circuits to dynamically field-programmable qubit arrays (DPQA).
        \textbf{(a)} Non-local connectivity of DPQA.
        Atoms are kept in traps generated by a 2D acousto-optic deflector (AOD, dashed grid) and a spatial light modulator (SLM, all others).
        Entangling two-qubit gates are enabled by a Rydberg laser illuminating the plane (glow).
        Only when two atoms are within the Rydberg blockade range $r_b$ can they perform an entangling gate (pairs in colored ovals).
        We can change the location of AOD atoms, and transfer atoms between AOD and SLM traps~\cite{natphys07-beugnon-tweezer} \textit{in the middle of computation} (each arrow corresponding to some AOD reconfiguration).
        Through such reconfigurations, \textit{new non-local} connectivities are established (oval dashes), i.e., different pairs of atoms can now perform entangling gates.
        \textbf{(b)} Our compilation approach.
        The input consists of the quantum circuit to execute and the DPQA architecture specification, e.g., how large the plane is and how many AOD rows and columns we can have.
        The compiled instructions have to respect the constraints of DPQA.
        For example, when a two-qubit gate is executed, the two qubits should be closer than $r_b$ and there cannot be another qubit nearby.
        Also, all traps in the same AOD row/column move together and must stay in the same order from the beginning to the end of the process.
        We formulate all the constraints to a satisfiability modulo theories (SMT) model and use an existing SMT solver to find solutions, with which we can derive valid DPQA instructions to run the circuit.
        \textbf{(c)} Structure of compiled results.
        We discretize space by prescribing \textit{interaction sites} shown as the proximity of integer points in the plane.
        The distance between sites is sufficient to suppress Rydberg interaction strengths~\cite{nature22-lukim-bluvstein-atom-array, evered2023highfidelity} so the two-qubit entangling gates can only take place within sites.
        Our compiler places the qubits in the quantum circuit to atoms in SLM or AOD at a specific interaction site in the beginning of execution.
        We discretize time by setting \textit{stages} when two-qubit gates are performed.
        After each stage, some AOD movements and atom transfers serve as routing for the gates executed at the next stage.
        }
    \label{fig:1}
\end{figure*}

\section{Introduction} \label{sec:intro}
The power of quantum computing relies on the ability to generate large-scale entanglement among qubits.
Entangling operations such as two-qubit gates requires qubits to interact, which often confines gate connectivity to be geometrically local.
Since superconducting quantum processors are fabricated on a 2D plane \cite{web21-google-sycamore-data-sheet, web22-ibm-quantum-processor, web22-rigetti-qpus}, the qubit connectivities are planar with a low node degree for practical reasons \cite{prx20-chamberland-zhu-yoder-hertzberg-cross-codes-low-degree}.
For small trapped-ion quantum processors \cite{web22-honeywell-h1, web22-ionq}, the connectivity is all-to-all.
However, it is challenging to maintain this feature when scaling up to multiple ion traps \cite{nature02-kielpinski-monroe-wineland-architecture-ion}, although exciting progress is being made~\cite{pino2021demonstration}.

Recently, neutral atoms trapped in arrays of optical tweezers have become a leading experimental platform for quantum computing. 
These systems are readily scaled to large numbers: Ebadi et al.~\cite{science22-lukin-mis} have operated up to 289 neutral atom qubits, and significant increases in system size are expected to continue.
Neutral atoms have recently also reached state-of-the-art fidelities: Evered et al.~\cite{evered2023highfidelity} realized parallel CZ gates on 60 qubits with fidelity $99.5\%$.
Moreover, Bluvstein et al.~\cite{nature22-lukim-bluvstein-atom-array} have demonstrated dynamically field-programmable qubit arrays (DPQA) where the qubit connectivity can be reconfigured dynamically \textit{during the computation itself}, as illustrated by \autoref{fig:1}a.
We focus on the DPQA architecture, aligning with the settings established in these experimental works. Specifically, the two-qubit gates are driven by a global Rydberg laser.

DPQA opens the field up to new opportunities for running quantum circuits with non-local connectivities and a high degree of parallelism.
However, in addition to its flexibility, there are hardware constraints, as shown in \autoref{fig:1}b.
In the compilation flow of quantum computing, \textit{layout synthesis} (Appendix~\ref{sec:mapping}) places the qubits and routes them to execute the gates at the appropriate time steps, as depicted in \autoref{fig:1}c.
Quantum layout synthesis has been studied for years \cite{dac23-olsq2, tc20-tan-cong-optimality-layout-queko, iccad20-tan-cong-optimal-layout-synthesis, asplos19-li-ding-xie-sabre-mapping, aspdac19-zulehner-wille-su4-compiling, dac19-wille-burgholzer-zulehner-mapping-minimal-swaph,  iccad19-bhattacharjee-saki-alam-chattopadhyay-ghosh-muqut-mapping,  isca19-murali-linke-martonisi-abhari-nguyen-alderete-triq-architecture-studies, asplos21-zhang-hayes-qiu-jin-chen-zhang-time-optimal-mapping, iccad21-tan-cong-qubit-mapping-absorption, tcad08-maslov-falconer-mosca-placement, aspdac14-shafaei-saeedi-pedram-placement-communication-2d, arxiv1703-Bhattacharjee-Chattopadhyay-depth-optimal-placement, cgo18-siraichi-santos-collange-pereira-qubit-allocation, dac19-ashsaki-alam-ghosh-qure-nisq, dac20-alam-ash-saki-ghosh-compilation-flow-qaoa, socs18-botea-kishimoto-marinescu-complexity-quantum-compilation, isca22-patel-silver-tiwari-geyser-neutral, isca21-baker-litteken-duckering-hoffmann-bernien-chong-long-distance-neutral, iccad21-brandhoher-buchler-polian-optimal-mapping-atoms}, but most previous works focus on fixed architectures.
A notable exception is Brandhofer et al.\cite{iccad21-brandhoher-buchler-polian-optimal-mapping-atoms}, which explores an architecture featuring only `1D displacements', much more restricted than DPQA.
Consequently, no previous compiler can fully leverage the reconfigurability and non-local connectivity of DPQA while conforming to all its constraints.
Realizing a compiler for DPQA is an outstanding challenge that would enable unique opportunities in quantum computing with such a flexible architecture.

In this work, we realize layout synthesis of complex quantum circuits for neutral atom hardware in a compiler OLSQ-DPQA (optimal layout synthesizer of quantum circuits for DPQA).
We encode states of the architecture in discrete variables specifying the location of qubits and the scheduling of gates.
Based on these variables, we express constraints of DPQA with first-order logic and integer relations.
Then, we use a satisfiability modulo theories (SMT) solver to derive valid variable assignments under the constraints yielding valid DPQA instructions to execute circuits.

The manuscript is organized as follows.
In Section~\ref{sec:description}, we review the DPQA architecture, especially its constraints.
In Section~\ref{sec:discrete}, we introduce the discrete variables that encode the state of DPQA in the computation spacetime.
Section~\ref{sec:optimal} explains how the constraints are constructed in SMT and how to invoke SMT solvers to derive optimal solutions with respect to circuit depth.
In Section~\ref{sec:greedy}, we introduce a hybrid method including a greedy heuristic to accelerate the compilation and scale it to large sizes.
Note that this method still relies on SMT, which can take exponential runtime in the number of qubits.
We demonstrate the effectiveness of our compiler by comparing the results of both optimal and optimal-greedy hybrid approaches to fixed planar architectures using state-of-the-art compilers.
Finally, in Section~\ref{sec:outlook}, we conclude the paper and provide future outlooks.

\section{DPQA Architecture Description} \label{sec:description}
In DPQA, every atom encodes a qubit, and they are held in optical traps.
A spatial light modulator (SLM) generates some of the traps.
These SLM-generated traps are stationary just like quantum registers in a fixed architecture, but can be placed in arbitrary locations~\cite{Browaeys_2016, browaeys_science_2016, Labuhn2016TunableTA}.
However, there are also \textit{mobile} traps~\cite{natphys07-beugnon-tweezer} generated by a crossed 2D acousto-optic deflector (AOD) are represented by the dashed grid in \autoref{fig:1}a~\cite{Scholl2021QuantumSO, nature21-keesling-256-atom-simulator}.
Atoms can be transferred between AOD and SLM.
If two atoms are within a certain \textit{Rydberg blockade range} $r_b$, we can apply an entangling two-qubit gate on the pair with a so-called Rydberg laser~\cite{Urban2008ObservationOR, prl19-lukin-parallel-gate-atom, evered2023highfidelity}.
In fact, the laser illuminates the whole plane where atoms are kept, so all pairs of qubits that are within distance $r_b$ perform two-qubit gates in parallel (pairs in colored ovals in \autoref{fig:1}a)~\cite{nature22-lukim-bluvstein-atom-array}. 

Different configurations, i.e., different locations of AOD rows and columns, lead to different qubit connectivities.
In principle, we can couple any AOD qubit $q$ at $(x_q,y_q)$ and any SLM qubit $q'$ at $(x_{q'},y_{q'})$ by moving $q$ to $q'$ by a distance roughly $(x_{q'}-x_q,y_{q'}-y_q)$.
Also, we can bring adjacent AOD rows/columns close together to couple qubits in these rows/columns.
The dashed ovals in \autoref{fig:1}a correspond to some of these \textit{potential} couplings that some AOD reconfigurations (arrows, not necessarily simultaneously) can achieve.

Although the DPQA architecture offers a lot of flexibility, it also has highly non-trivial constraints.
As mentioned previously, two qubits need to be closer than $r_b$ for an entangling gate, but if there is another qubit not sufficiently separated ($<2.5~r_b$) from the pair, the quantum evolution on these three qubits would not be a desired gate (`interaction exactness' constraint in \autoref{fig:1}b).
The 2D AOD grid is the product of two 1D AODs as the X and Y components.
There is one AOD trap at every intersection of the X and Y components.
Thus, we cannot move AOD traps individually.
What we can do to reconfigure the qubit connectivity is shift whole rows and/or columns of AOD traps.
Moreover, AOD rows/columns are not allowed to move across other rows/columns (to avoid heating / loss of the atoms during this process)~\cite{nature22-lukim-bluvstein-atom-array}, so the order of AOD rows/columns cannot change.
For a more detailed and formal discussion on DPQA, please refer to Appendix~\ref{sec:physics}.

\section{Discretization of the Solution Space} \label{sec:discrete}

\begin{figure*}
    \centering
        \includegraphics[width=\linewidth]{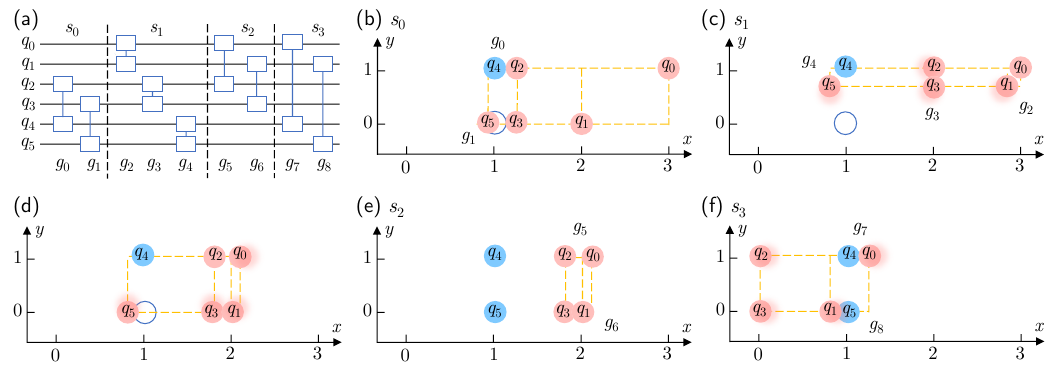}
        \caption{A compiled example.
        \textbf{(a)} The quantum circuit to compile.
        \textbf{(b)} Stage 0.
        Qubits are loaded to the corresponding traps before this stage: blue qubits are in SLM, red qubits are in AOD.
        An AOD trap sits at every intersection of the AOD columns and rows (x and y dashed lines).
        An open circle represents an unoccupied SLM trap.
        At stage 0, $(q_4, q_2)$ and $(q_5, q_3)$ are at same sites to enable a Rydberg interaction.
        Thus, two gates $g_0$ and $g_1$ are applied to these two pairs of qubits.
        After stage 0, the movement shifts the lower AOD row from $y=0$ to $1$ and the middle two columns go from $x=1$ and $2$ to $x=2$ and $3$, respectively.
        \textbf{(c)} Stage 1.
        Shadows of qubits indicate the direction of the movements from the previous stage to the current one.
        \textbf{(d)} The moment after the movement between stage 1 and 2.
        \textbf{(e)} Stage 2.
        $q_5$ is transferred from AOD to SLM (red to blue) after the movement and before stage 2 by shifting the leftmost AOD column to align with the SLM trap at $(1,0)$ and then turning off this column.
        \textbf{(f)} Stage 3 finishing the circuit execution.
}
    \label{fig:2}
\end{figure*}

As pointed out previously, we have the freedom to specify the locations of an AOD row $r$ as a function of time $y_r(t)$ and, similarly, $x_c(t)$ of an AOD column $c$.
Modeling the DPQA architecture based on these continuous functions is cumbersome and unnecessary for a compiler.
In fact, the time domain can be easily discretized to \textit{stages} like in \autoref{fig:1}c because we only care about the location of qubits when we turn on the Rydberg laser to apply the entangling gates.
As long as we do not violate the DPQA constraints, the 2D planar movements of AOD between any two stages can be straightforwardly interpolated.
We can implement single-qubit gates using individually addressable lasers between stages, so we filter out the two-qubit gates and compile them.
After this compilation, we can reintroduce single-qubit gates.
For more details, kindly refer to Appendix~\ref{ssec:generic}.

We also discretize the space domain to \textit{interaction sites}.
The intuition behind the spatial discretization is that the spatial sparsity of qubits is required to avoid unwanted Rydberg interaction.
At each stage, the qubits need to be either in a pair sitting close to each other (performing two-qubit gates) and away from all other qubits or all alone (idling) and away from all other qubits.
Thus, we restrict the location of qubits at each stage to the proximity of pre-defined grid points (interaction sites) on the 2D plane, as shown in \autoref{fig:1}c.
The unit separation between these sites is sufficiently large so that no Rydberg interaction can take place among qubits at different sites.

With the above discretizations, we can use discrete variables to encode all possible configurations of the architecture.
We shall work through the example of compiling the quantum circuit in \autoref{fig:2}a to DPQA to explain the variables.
More details about spatial discretization are provided in Appendix~\ref{ssec:discrete-space}.
All values in the running example are provided in Appendix~\ref{ssec:json}.

\textit{Site indices $x_{i,s}$ and $y_{i,s}$:} at stage $s$, qubit $q_i$ is at interaction site $(x_{i,s}, y_{i,s})$, e.g, for $q_0$ at stage 0 (\autoref{fig:2}b), $x_{0,0}=3$ and $y_{0,0}=1$; at stage 1 (\autoref{fig:2}c), $x_{0,1}=3$ and $y_{0,1}=1$ still; at stage 2 (\autoref{fig:2}e), $y_{0,2}=1$ still, but $x_{0,2}=2$ due to movements.

\textit{Array index $a_{i,s}$:} at stage $s$ and the movement following it, if $q_i$ is in SLM, $a_{i,s}=0$; if it is in AOD, $a_{i,s}=1$, e.g., for $q_5$ at stages 0 and 1, $a_{5,0}=a_{5,1}=1$; before stage 2, it is transferred to SLM, so $a_{5,2}=0$.

\textit{AOD column/row indices $c_{i,s}$ and $r_{i,s}$:} at stage $s$ and the movement following it, qubit $q_i$ is at AOD column $c_{i,s}$ and row $r_{i,s}$, e.g., at stage 0, $r_{5,0}=0$ and $c_{5,0}=0$ for $q_5$; $r_{0,0}=1$ and $c_{0,0}=3$ for $q_0$.
(We index the row from below and the column from left.)
Since it is unknown in advance whether a qubit will be in AOD or SLM, we introduce the $r$ and $c$ variables for all qubits, but only those for AOD qubits will play a role in constraints.

\textit{Time coordinate $t_j$:} gate $g_j$ is scheduled to stage $t_j$, e.g., $g_0$ in \autoref{fig:2}a is on $q_2$ and $q_4$ and at stage 0, so $t_0=0$; $g_1$ is also at stage 0 (\autoref{fig:2}b), so $t_1=0$.
$g_7$ and $g_8$ are at stage 3 (\autoref{fig:2}f), so $t_7=t_8=3$.

\begin{figure}[p]
    \centering
        \includegraphics[width=\linewidth]{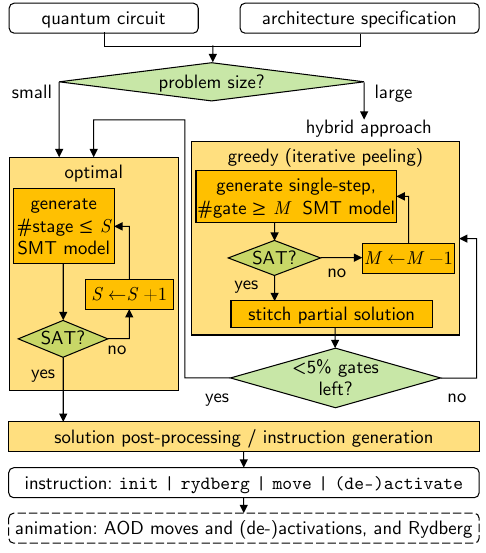}
        \caption{Workflow of our compiler OLSQ-DPQA.
        The inputs to the compiler are the quantum circuit to execute and the specifications of the DPQA architecture considered.
        If the problem is small, the compiler directly takes the optimal approach by constructing an SMT model where all the gates are applied to the first $S$ stages.
        If the model is satisfiable, then we find a solution; otherwise, we increase $S$ and try again.
        Thus, we find a solution with the minimum number of stages in the end, because lower-depth models are all checked and  unsatisfiable.
        The SMT solution goes through a post-processing to extract the instructions for executing the quantum circuit on DPQA.
        There are only five types of instruction: \texttt{init} for initialization; \texttt{rydberg} to turn on the Rydberg laser and perform two-qubit gates; \texttt{move} for changing the coordinates of AOD rows/columns; \texttt{activate} for turning on certain AOD rows/columns for atom transfer; and \texttt{deactivate} for turning off certain AOD rows/columns.
        If the problem is large, the compiler takes a hybrid approach by iteratively ``peeling off'' the maximum number of gates possible.
        It generates a single-step (two-stage) SMT model with a constraint of executing more than $M$ gates in one step.
        After possible decreases of $M$, we find the solution with as many gates executed in one step as possible.
        Then, we stitch this partial solution, which is one ``layer peeled off'', to the whole solution.
        When the problem becomes sufficiently small ($5\%$ of gates left), the compiler switches to the optimal approach.
        }
    \label{fig:3}
\end{figure}

\section{Optimal Compilation with SMT} \label{sec:optimal}
Given the variables above, we can express the DPQA constraints.
Usually, the constraints can be expressed with integer inequality and first-order logic.
The simplest ones are the bounds for variables.
As a part of the architecture specification, we restrict the region where qubits can be put and moved.
Depending on our field of view and $r_b$, there are upper bounds $X$ and $Y$ of how many traps we have in horizontal and vertical directions. 
So, we have integer inequality constraints:
\begin{equation}
    0 \le x_{i,s} < X,\ 0 \le y_{i,s} < Y \quad \forall i, s.
\end{equation}
Depending on how many AOD columns and rows are at our disposal, we also restrict the range of $c_{i,s}$ and $r_{i,s}$ with constants $C$ and $R$.
Other constraints may also require Boolean logic, e.g., the qubits in SLM are stationary between two stages:
\begin{equation}
    \left(a_{i,s}=0\right) \Rightarrow\left(x_{i,s+1}=x_{i,s}\right)\wedge\left(y_{i,s+1}=y_{i,s}\right) \quad \forall i,s.
\end{equation}
For example, $q_4$ is in SLM at stage 0, i.e., $a_{4,0}=0$, so its site indices remain the same between stage 0 and 1, i.e., $x_{4,1}=x_{4,0}$ and $y_{4,1}=y_{4,0}$.

For a valid two-qubit gate, enforcing \textit{connectivity} constraints is crucial.
Taking $g_4$ as an example, with $q_4$ and $q_5$ requiring the same site, we express the constraint as
\begin{equation}
(t_4 = 1)\Rightarrow\ ((x_{4,1}=x_{5,1}) \wedge (y_{4,1}=y_{5,1})).
\end{equation}
AOD qubits, shifting by whole rows or columns during movement, maintain constant row or column indices across consecutive stages.
For instance, with $q_5$ in AOD at stage 0, we have:
\begin{equation}
(a_{5,0}=1) \Rightarrow ((c_{5,1}=c_{5,0}) \wedge (r_{5,1}=r_{5,0})).
\end{equation}
To uphold order constraints during movement, we enforce ordering of the site indices in the next stage.
For instance, at stage 0, where $q_1$ is at row 0 and $q_0$ is at row 1, the constraint is expressed as
\begin{equation} \label{eq:eq-eg}
((a_{0,0}=1) \wedge (a_{1,0}=1) \wedge (r_{1,0}<r_{0,0}))\Rightarrow (y_{1,1} \le y_{0, 1}).
\end{equation}
Following the above approach, all DPQA constraints can be formulated using first-order logic and integer inequalities, facilitating automatic reasoning.
Interested readers can refer to the full list in Appendix~\ref{sec:formulation}.

Once acquainted with the formulation, adapting constraints to accommodate architectural changes is straightforward.
For instance, if the Rydberg laser only illuminates the left part of the plane up to $X_L$, the connectivity constraint can be revised as follows:
\begin{equation}
(t_4 = 1)\Rightarrow\ ((x_{4,1}=x_{5,1}) \wedge (y_{4,1}=y_{5,1}) \wedge (x_{4,1}\le X_L)).
\end{equation}
If there are multiple AODs, the array index variables can be generalized to have a larger support than just 0 and 1.
For instance, in addition to \autoref{eq:eq-eg}, an additional constraint is introduced for a second AOD array:
\begin{equation}
((a_{0,0}=2) \wedge (a_{1,0}=2) \wedge (r_{1,0}<r_{0,0}))\Rightarrow (y_{1,1} \le y_{0, 1}).
\end{equation}
Employing this approach, one can tailor the constraints to specific architectural settings.

\begin{figure*}
    \centering
        \includegraphics[width=\linewidth]{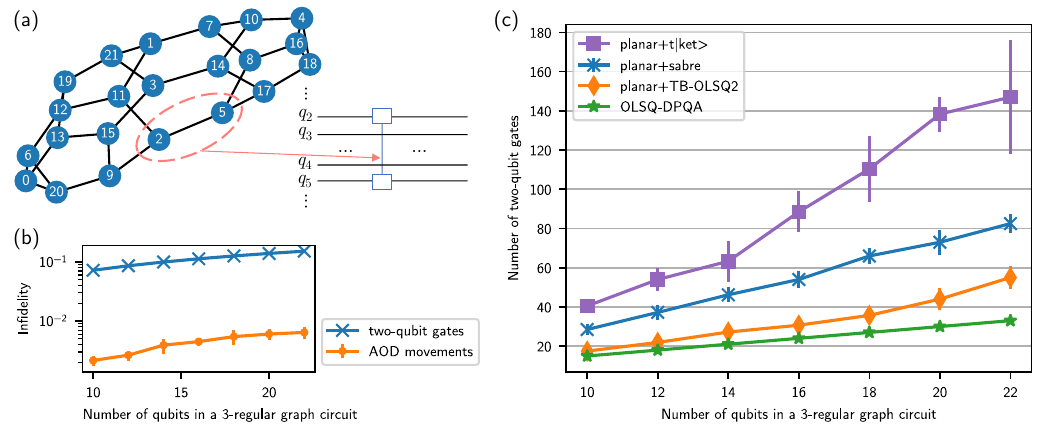}
        \caption{Evaluation of the optimal compiler.
        \textbf{(a)} Graph circuits.
        Given any graph, we treat each node as a qubit and add a two-qubit entangling gate for every edge in the graph to construct the graph circuit.
        We assume the gates are commutable, so gate order does not matter.
        The benchmarks used are graph circuits generated by 3-regular graphs of size 10 to 22.
        For each size, we have 10 random graphs.
        \textbf{(b)} Comparison of infidelity caused by the Rydberg laser (performing two-qubit gates) and the AOD movements.
        The latter is 27x smaller on average.
        We make such estimation using $99.5\%$ two-qubit gate fidelity~\cite{evered2023highfidelity} and a movement scheme that yields low atom heating as in Ref.~\cite{nature22-lukim-bluvstein-atom-array}.
        \textbf{(c)} Comparison of the number of two-qubit gates required on a fixed planar architecture (Google's Sycamore) and DPQA employing different compilers.
        Error bars are standard deviations among 10 random graphs of the same size.
        The compilers are t$|$ket$\rangle$, SABRE (integrated in Qiskit), and TB-OLSQ2.
        Note that TB-OLSQ2 is optimal for fixed architectures, but there is still a significant gap (1.7x) between it and the optimal DPQA compiler.
        The gaps mainly come from SWAP gates inserted on the fixed architecture, which requires three entangling gates (controlled $R_z$)~\cite{micro20-gokhale-javadi-abhari-earnest-shi-chong-compilation-openpulse}.}
    \label{fig:4}
\end{figure*}

Satisfiability modulo theories (SMT) solvers are tools that can solve valid variable assignments given this form of constraint.
SMT is an extension of Satisfiability (SAT) that accommodates a broader range of variable types beyond binary variables, as well as diverse types of constraints that go beyond the confines of conjunctive normal form.
We can encapsulate the variable definitions and constraints expressed with these variables in an SMT \textit{model}.
When provided with a model, an SMT solver can check whether it is satisfiable.
If so, the solver returns the variable assignments which are all we need to execute circuits because our variables completely capture the architecture in spacetime.
If the model is not satisfiable, some of our bounds are too small for valid variable assignments that will satisfy all the constraints.

With an SMT solver, we are able to not only solve valid assignments to compile circuits, but also guarantee the optimality of the solution with respect to some objective function, presented as the optimal branch in \autoref{fig:3}.
So far, the dominating error source of DPQA is the Rydberg laser (see Appendix~\ref{ssec:error} for detail), so we ignore the single-qubit gates for now and add them back after layout synthesis.
Then, our objective is the number of times we turn on the Rydberg laser, i.e., the number of stages of parallel two-qubit gates, or, circuit depth $S$.
Thus, we use relatively large spatial bounds ($X,Y,R,C$) which is more likely to yield satisfiable models with a lower $S$.
We start by setting $S$ to a lower bound, e.g., the critical path in the circuit, which is 3 for the one in \autoref{fig:2}a.
If the SMT solver returns unsatisfiable, we increase $S$ and invoke the solver again, until it finds a valid solution with $S_\mathrm{opt}$ stages.
With this procedure, the optimality is guaranteed since we have checked for any smaller $S$ yield unsatisfiable models before finding the solution.
If $S$ increases beyond the number of gates, we conclude that the spatial bounds are too small and increase them.
The procedure will terminate since any finite circuit of size $P$ can be run in a finite spacetime volume bounded by $P\times P \times P$.

The variable assignments are not yet a DPQA executable.
We need to post-process the SMT solution to produce DPQA instructions.
For example, we must know the beginning and end coordinates of each AOD column, which are stored distributively in the $x_{i,s}$ and $c_{i,s}$ variables.
In the example of \autoref{fig:2}, we find $q_2$ is in column $1$ and $x=1$ at stage 0, i.e., $c_{2,0}=1$ and $x_{2,0}=1$, and $x=2$ at stage 1, i.e., when $x_{2,1}=2$, we infer that the AOD column 1 travels from $x=1$ to $x=2$.
As such, the information in the SMT solution will be translated to five types of basic DPQA instructions: \texttt{init} for initial qubit loading, \texttt{rydberg} for illuminating the Rydberg laser, \texttt{move} for AOD movements, \texttt{activate} for activating AOD rows/columns, and \texttt{deactivate} for deactivating AOD rows/columns.
These instructions are readily executable on DPQA and our compiler can also generate animations from the instructions to view the execution process in action.

We benchmark the effectiveness of DPQA and our compiler, OLSQ-DPQA, on a set of quantum circuits constructed using random graphs, as illustrated in \autoref{fig:4}a.
For a given graph, we assign each node to a qubit and apply a two-qubit gate for every edge.
For simplicity, we consider problems where these gates are commutable, like the controlled-$R_z$ gates available on DPQA~\cite{prl19-lukin-parallel-gate-atom, evered2023highfidelity}, so the compiler also explores freedom of permuting gate ordering.
Compiling these circuits is more challenging compared to generic circuits due to the increased flexibility in commutation.
For evaluations on realistic generic circuits, please refer to Appendix~\ref{ssec:generic}.

By considering measured 1.5~s coherence times~\cite{nature22-lukim-bluvstein-atom-array} and 99.5\% CZ gate fidelity~\cite{evered2023highfidelity} in DPQA, as expected, we find the main error source is the Rydberg laser (\autoref{fig:4}b).
The infidelity associated with AOD movements based on real experimental parameters (details in Appendices~\ref{sec:physics} and \ref{sec:evaluation}), is revealed to be $\sim 27\times$ smaller than the infidelity of two-qubit gates, as shown in \autoref{fig:4}c, indicating that DPQA is promising for realizing highly nonlocal graphs where motion can be complex, and that increasing two-qubit gate fidelity is the main way to continue boosting fidelity.

We now compare our DPQA compilation results to the compilation results on fixed planar architectures, where instead of physically moving qubits around, qubits are moved around using two-qubit SWAP gates. As expected, DPQA combined with the optimal compiler requires significantly fewer two-qubit gates.
We tested a few compilers that perform layout synthesis for the fixed architecture by inserting SWAP gates: 
t$|$ket$\rangle$~\cite{qst20-sivarajah-dikes-cowtan-simmons-edgington-duncan-tket-compiler-nisq}, a heuristic compiler used in leading QAOA experiments \cite{natphys21-google-qaoa};
SABRE~\cite{asplos19-li-ding-xie-sabre-mapping}, a heuristic compiler integrated in leading quantum programming framework, Qiskit~\cite{Qiskit};
and TB-OLSQ2~\cite{dac23-olsq2}, a leading optimal compiler for fixed architecture.
The gaps of the number of two-qubit gates for a QAOA benchmark with 22 nodes in the graph are 4.5x, 2.5x, and 1.7x, respectively.

\section{Hybrid Approach} \label{sec:greedy}

\begin{figure*}
    \centering
        \includegraphics[width=\linewidth]{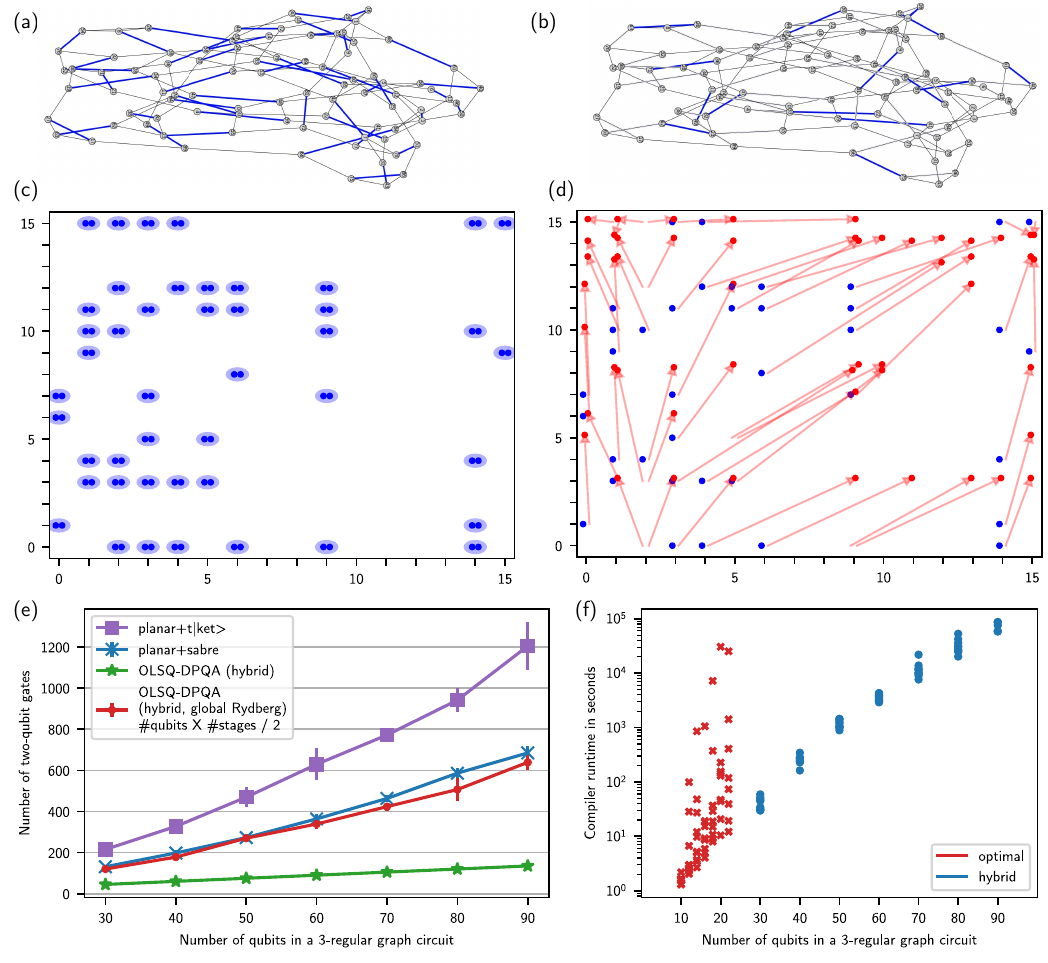}
        \caption{Evaluation of the greedy-optimal hybrid compiler.
        \textbf{(a, b)} One of the largest benchmarks we are able to compile, a 90-node 3-regular graph.
        The highlighted edges are gates executed at the stages in (c) and (d), respectively.
        \textbf{(c)} One stage of the compiled result.
        The dots are qubits in SLM.
        The ovals indicate two-qubit gates performed at this stage, which have a 1-to-1 correspondence with the edges in (a).
        After this stage, some qubits are transferred to AOD and moved.
        \textbf{(d)} The next stage.
        The red dots are the AOD qubits, and the arrows indicate the parallel movements from (c) to the current state.
        Readers are welcome to check out our code base for this animation.
        \textbf{(e)} Comparison of the number of two-qubit gates required on a fixed planar architecture (10x10 grid) using different compilers and DPQA.
        For DPQA, the number of two-qubit gates scales as $n$, whereas for the state-of-the-art heuristic solver on the fixed planar architecture, SABRE, scales as $n^{1.52\pm0.02}$ where $n$ is  the number of qubits.
        DPQA requires far less two-qubit gates, 5.1X less than SABRE, and scales linearly.
        \textbf{(f)} Comparison of runtime of the optimal and hybrid approaches in OLSQ-DPQA.
        Since both of them internally rely on SMT solving, the runtime scalings are both exponential in the size of the graph with which we generate the quantum circuit.
        However, the hybrid approach is significantly faster so that large instances can be solved (up to 90 qubits in $10^5$ $\sim$ a day).
        Compared to the optimal approach, the scaling of the hybrid approach is mainly related to size rather than the specific graph, which is demonstrated by the much smaller spread of data points at the each size.}
    \label{fig:5}
\end{figure*}

The runtime for SMT solving scales exponentially in the worst case, so the optimal compiler can take a very long time to solve certain cases, as seen in \autoref{fig:5}f.
Due to the complicated constraints, it is also challenging to design near-optimal purely heuristic algorithms to search the solution space of DPQA.
Therefore, we adopt a two-level approach, as illustrated in the hybrid approach in \autoref{fig:3}.
For large problems, at the higher level, we apply a greedy heuristic in that, at every stage, we find the AOD movement to maximize the number of gates to execute in the next stage and we repeat this process until there are a sufficiently small number of gates remaining.
Then, we switch to the optimal approach.
This technique is inspired by `iterative peeling' for routing classical integrated circuits with multi-layer routing layers \cite{cong_provably_1993}.

Specifically, if there are still gates to execute, we construct a ``single-step'' SMT model with two stages and set the qubit location of the initial stage to that of the current stage in the full solution.
For instance, suppose the compiler has already processed $g_0$ to $g_4$ in \autoref{fig:2}a and progressed to stage 1 (\autoref{fig:2}c).
In the single-step model, all initial locations are set by the previous partial solution, e.g., $x_{4,0}=1$ since $q_4$ at $x=1$ in \autoref{fig:2}c.
Then, we optimize the number of gates executed in the second stage in the single-step model with a procedure similar to the one in the optimal approach, except that we decrease the number of gates to execute in the each stage from an upper bound of $M$ instead of increasing the number of stages from a lower bound. 
The upper bound is the maximum matching number of the graph constructed from the remaining gates.
In our example, the remaining gates $g_5$ and $g_7$ both act on $q_0$, whereas $g_6$ and $g_8$ both act on $q_1$, so only one gate in each of these two pairs can be executed together, i.e., the maximum matching number is 2.
The compiler appends the single-step model with a constraint that says there are at least 2 gates executed at the next stage and invokes the SMT solver, which can find such a solution (\autoref{fig:2}e).
We stitch this partial solution to the full solution, remove gates $g_5$ and $g_6$ which is a layer of gates ``peeled off'', and continue to the next ``peeling''.
If there are only a few gates remaining (we opt for $5\%$) the compiler switches to the optimal approach to solve for the final stages.

The hybrid approach cannot fundamentally improve the runtime scaling to polynomial because it still relies on SMT solving, but it greatly accelerates the process, with some sacrifice on optimality.
As exhibited in \autoref{fig:5}f, it is much faster than the optimal compiler and the divergence of runtime within benchmarks of the same size is also much smaller.
Within a reasonable amount of time ($10^5$s $\approx$ a day), the hybrid compiler managed to compile some 90-qubit circuits whereas the optimal compiler, in the worst case, can only compile up to the 22-qubit circuit.
We present one of the largest circuits compiled in \autoref{fig:5}a-d: (a, b) exhibit the graph generating the quantum circuit which has a complex connectivity;
(c, d) are two stages in the program execution.
This hybrid approach is implemented in the OLSQ-DPQA compiler, which is open-source under the BSD 3-clause license.\footnote{\url{https://github.com/UCLA-VAST/DPQA}}
The code base includes Python scripts that 1) generate the SMT models and iteratively invoke an SMT solver, Z3~\cite{tacas08-demoura-bjorner-z3-smt-solver} to solve them, 2) generate DPQA instructions and animations based on SMT solutions, 3) draw plots in the evaluations.
The dependencies are Python packages Z3~\cite{tacas08-demoura-bjorner-z3-smt-solver}, PySAT~\cite{imms-sat18}, NetworkX~\cite{lanl08-hagberg-swart-chult-networkx}, and Matplotlib~\cite{matplotlib-Hunter:2007}.
The code base also includes all SMT solutions in the evaluations and some example animations.

We compare the required number of two-qubit gates by DPQA and a fixed planar architecture (10x10 grid) in \autoref{fig:5}e.
We find that the savings from DPQA on such a large system is significant compared to the fixed architecture: 5.1x and 8.9x reduction in the number of two-qubit gates, respectively, compared to the compilation results by SABRE and t$|$ket$\rangle$.
If the heuristics place qubits in an $\sqrt{n}$-by-$\sqrt{n}$ region, each gate may require $O(\sqrt{n})$ SWAPs to route.
Then, for $O(n)$ gates, as in our benchmark set, $O(n^{1.5})$ SWAPs are required.
We observe this scaling in the results of SABRE: with a log-log fitting, the number of two-qubit gates scales in the $1.52\pm 0.02$ power of the number of qubits.
In comparison, DPQA routes the gates by AOD movements instead of SWAPs, so the number of gates scales linearly.
This result assumes that DPQA is equipped with an individually addressable Rydberg laser (or other methods of turning off the Rydberg excitation locally) that does not accumulate the same error on idling qubits (e.g., in \cite{nature22-graham-atom-array}).\footnote{Instead, if DPQA is equipped only with a global Rydberg laser that illuminates the whole plane, although the number of two-qubit gates is greatly reduced, the effective number of two-qubit gates (the number of qubits times the number of stages divided by 2) is only slightly better ($7\%$) than the SABRE results on the fixed architecture \textit{assuming that} a global Rydberg laser induces the similar error rate, at every stage, on idling qubits as well as qubits involved in two-qubit gates~\cite{nature22-lukim-bluvstein-atom-array}.
}

\section{Discussion and Outlook} \label{sec:outlook}

In this work, we studied the constraints of existing dynamically field-programmable qubit arrays (DPQA) architecture regarding compilation and discretization of the architecture.
Then, we developed a compiler for the architecture, OLSQ-DPQA.
The compiler and results developed here are suited to a wide range of applications in neutral atom quantum computation.
Broadly, combined with state-of-the-art neutral atom hardware and the ability of our compiler to output hardware-level instructions, these techniques would allow us to specify an arbitrary quantum circuit with arbitrary connectivity on approximately 100 qubits and then realize this circuit on quantum hardware.
This can allow, for example, implementing circuits like the quantum fourier transform on significantly larger numbers of qubits than previously realized~\cite{weinstein2001implementation, debnath2016demonstration}, implementing nonlocally connected, high-rate quantum low-density parity check (LDPC) codes~\cite{quantum21-decoders-ldpc}, or studying evolution in exotic Hamiltonians~\cite{kalinowski2023nonabelian}.

In particular, the results presented for realizing circuits on random 3-regular graphs on 90 qubits can directly be applied to various problems.
For combinatorial optimization, one can utilize the class of graphs examined here to study the quantum algorithm performance on problems such as MAXCUT or MIS with either the Quantum Adiabatic Optimization Algorithm (QAOA)~\cite{farhi2000quantum} or the trotterized adiabatic evolution~\cite{science22-lukin-mis}.
Realizing such a complex evolution on nonlocally connected graphs such as in \autoref{fig:5} can be used for efficient quantum supremacy and information scrambling experiments~\cite{nature19-google-quantum-supremacy, optical-quantum-supremacy}.
In such an approach, one can choose to employ the first few, most efficient layers of the compilation to implement classes of random nonlocally connected graphs.  

The compiler used here can be expanded further along multiple axes.
Given the generality and flexibility of the framework, the compiler can adapt to new hardware features before implementation and inform hardware design for neutral atom quantum computers. 
We demonstrated a significant gate reduction showcasing the power of the reconfigurable quantum computing architectures.
Such results incentivize further developments of DPQA hardware, e.g., scaling up DPQA to over 1000 qubits, including mid-circuit readout and quantum error correction, to execute large scale quantum circuits.
To support hardware at such a scale, a high-performance efficient compilation is required, and our formulation provides a solid basis to start constructing such compilation methods.
In particular, the SMT variables can function as the state variables within a tree search node or a machine learning agent, while the SMT constraints characterize the transitions to other states.
In addition, the comparison of results with local and global Rydberg laser control indicates the importance of hardware to have locally switchable Rydberg excitation~\cite{nature22-graham-atom-array} or ``idle'' zones~\cite{bluvstein2023logical}.
Qubits can be stored in these zones to avoid accumulating errors when other qubits are being operated on. 
The idling zones also simplifies the compilation problem since the qubits in these zones do not need to be spatially separated to avoid Rydberg interaction, which may greatly accelerate the compilation.

\section*{Acknowledgements}
This work is funded by NSF grant 2313083. Furthermore, it receives partial support from DOE (Quantum System Accelarator Center contract number 7568717), the Army Research Office MURI (grant number W911NF-20-1-0082), DARPA ONISQ program, and contributions from multiple industrial sponsors including NEC and Amazon Web Services.
The authors would like to thank Wan-Hsuan Lin for discussions on t$|$ket$\rangle$, SABRE, and TB-OLSQ2, Jennifer Yu-Hsuan Chen for revising part of the code base, Maddie Cain, Sepehr Ebadi, Simon Evered, Tom Manovitz, Giulia Semeghini, Tout Wang, and Harry Zhou for discussions on atom arrays. D.B. acknowledges support from the NSF Graduate Research Fellowship Program (grant DGE1745303) and The Fannie and John Hertz Foundation.

\bibliographystyle{quantum}
\bibliography{refs}

\appendix

\section{DPQA Architecture}\label{sec:physics}

In an atom array system, each individual qubit is a single atom trapped in an individual optical tweezer, which enables a deterministic control over the qubit position.
The physics of atomic trapping, optical tweezers, and entangling gates leads to several key implications.
These implications serve as the interface between physics and computer science with which we reason about the variables, constraints, and optimization procedure in our compilers.
Thus, we enumerate them in this appendix for reference.
For the specific technical parameters, we follow the state-of-the-art experimental work~\cite{evered2023highfidelity, nature22-lukim-bluvstein-atom-array}.

\subsection{Atom Trapping}

One trap cannot hold more than one atom.
Otherwise, the atoms may expel each other out of the trap.
\begin{implication} \label{impl:qubit-trap-mapping}
One trap can hold zero or one atom at any time during the computation.
\end{implication}

Two orthogonal optical components generate AOD tweezers.
The X component produces a horizontal pattern, and the Y component multiplies this pattern by a vertical pattern.
In contrast, an arbitrary phase hologram on a spatial light modulator produces SLM tweezers.
As a result, we can place each SLM tweezer in an arbitrary location.
However, to enable massive parallelism of gate execution, the geometry of the SLM and the AOD should be similar.
\begin{implication} \label{impl:optical-tweezer-array}
AOD and SLM optical trap arrays are rectangular arrays that extend in the X and Y direction in the 2D plane.
\end{implication} 
\noindent E.g., in \autoref{fig:2}b, the AOD is a rectangular array with two rows and four columns, indicated by the dashed grid.
The dynamically programmable processor in \cite{nature22-lukim-bluvstein-atom-array} uses up to 24 qubits, but system sizes of 100s of qubits are attainable as was done in \cite{science22-lukin-mis}, and both SLM and AOD grids have been used in system sizes as large as 16x16 each \cite{prx22-singh-dual-element}.

Because of the finite optical resolution of the microscope generating tweezers, traps of the same array cannot be closer than a given minimum spacing.
In \cite{nature22-lukim-bluvstein-atom-array}, it is $2\upmu$m.
\begin{implication} \label{impl:row-separation}
There is a minimal separation between two rows or columns of traps in the same array, $d_s$.
\end{implication}

\subsection{Array Movements} \label{ssec:movement}

AOD traps can move whereas SLM traps cannot.
Thus, it may seem to some readers that SLM is strictly less general than AOD, rendering the notion of SLM redundant for compilation.
However, an advantage of SLM is that we can turn off the unused traps based on the compilation result.
As part of the architecture specification, we make a certain number of SLM traps available to the compiler, but some of them are never used throughout the compiled result.
Then, we simply ignore them when we generate the SLM in the beginning of the experiment, which saves some laser power.
Although total laser power is not a bottleneck at the moment, the savings of SLM is beneficial for future scaling-up.
Thus, we keep SLM in the formulation instead of just treating it as a special case of AOD.
\begin{implication} \label{impl:slm}
If the array is the SLM type, the traps are stationary.
\end{implication}
\noindent E.g., $q_4$ stays at the same place throughout \autoref{fig:2}b-\ref{fig:2}f.

The control we have on the AOD traps are the Y coordinate of each row and the X coordinate of each column.
\begin{implication} \label{impl:aod-together}
If the array is the AOD type, a row/column of traps move together.
\end{implication}
\noindent E.g., from \autoref{fig:2}b to \ref{fig:2}c, the AOD row of $q_5$, $q_3$, and $q_1$ moves upwards, and the column of $q_2$ and $q_3$ moves to the right.

Per \autoref{impl:row-separation}, we cannot place two rows/columns too close together.
If rows A and B move across each other, they must have been closer than the minimum spacing at some point, which is prohibited.
\begin{implication} \label{impl:aod-order}
If the array is the AOD type, a row cannot cross over another row, a column cannot move over another column.
\end{implication}

In \cite{nature22-lukim-bluvstein-atom-array}, the relation between movement time $t$ and travel distance $D$ is set as $t=T_0\sqrt{D/D_0}$ to maintain constant heating of the atoms during movements.
We follow their setting $T_0=200\upmu$s and $D_0=110\upmu$m so that the heating is sufficiently low.

\subsection{Quantum Gates}

Single-qubit gates are high-fidelity operations that are generically easy to perform locally (see \cite{prl19-lukin-parallel-gate-atom}).
\begin{implication} \label{impl:single-qubit}
Arbitrary single-qubit gates can be addressed to each qubit individually.
\end{implication}

We perform two-qubit operations with a specific type of laser to excite the atoms to Rydberg state.
In this state, atoms within a certain distance will interact strongly and cannot be excited simultaneously.
The characteristic distance of this interaction is the Rydberg blockade radius $r_b$ ($7.5\upmu$m in \cite{nature22-lukim-bluvstein-atom-array}).
This blockade mechanism is the basis of two-qubit entangling gates; only if two atoms are within an $r_b$ of each other can they perform a two-qubit gate.
The specific gate implemented in \cite{nature22-lukim-bluvstein-atom-array} is the Levine-Pichler CZ gate, which is a special case of controlled-$R_z$ gates available in DPQA~\cite{prl19-lukin-parallel-gate-atom}.
\begin{implication} \label{impl:two-qubit-close-necessary}
Two qubits $q$ and $q'$ can only perform an entangling two-qubit gate when they are within a blockade radius, i.e., $|\vec{x}_q - \vec{x}_{q'} | \le r_b$, and they are both illuminated by the Rydberg laser.
\end{implication}

The Rydberg laser is \textit{global} in the sense that it illuminates \textit{all the qubits}, as done in \cite{science22-lukin-mis, nature22-lukim-bluvstein-atom-array}.
When we turn on the laser, we cannot ``switch off'' the interaction of a pair if they are within range.
\begin{implication} \label{impl:two-qubit-close-sufficient}
If $q$ and $q'$ are within $r_b$ and illuminated by the Rydberg laser, they will go through a two-qubit entangling gate.
\end{implication}

If two atoms are sufficiently separated, $>2.5r_b$ in practice, they will not interact even if excited by the Rydberg laser.
If there are more than two atoms that are not sufficiently separated, they go through a joint quantum process which is not a well-defined gate.
\begin{implication} \label{impl:two-qubit-no-third}
For any three qubits $q_0$, $q_1$, and $q_2$, at most one of the following is true when the Rydberg laser is on: $|\vec{x}_{q_0}-\vec{x}_{q_1}|<2.5r_b$, $|\vec{x}_{q_1}-\vec{x}_{q_2}|<2.5r_b$, $|\vec{x}_{q_2}-\vec{x}_{q_0}|<2.5r_b$.
That is, only disjoint pairs of qubits may entangle simultaneously.
\end{implication}

\subsection{Error Source} \label{ssec:error}

Errors can occur during the gates or the idling time (including AOD movements, activation, and deactivation).
In the evaluation in this paper, the average idling time is only $2.6\%$ of the qubit lifetime (coherence time) in the largest benchmark (90-node QAOA).
Thus, the idling time plays a relatively small role in the error source.
In addition, the single-qubit operations are significantly higher fidelity ($99.99\%$) than the two-qubit entangling gates ($99.5\%$) \cite{nature22-lukim-bluvstein-atom-array, evered2023highfidelity}.
A global Rydberg laser for the two-qubit gates induces the same error rate on all qubits whether they are involved with a two-qubit gate at this stage or not.
\begin{implication} \label{impl:main-error-source}
The main computational error source is the number of layers of two-qubit gates.
\end{implication}

\begin{figure}[t]
    \centering
    \includegraphics[width=\linewidth]{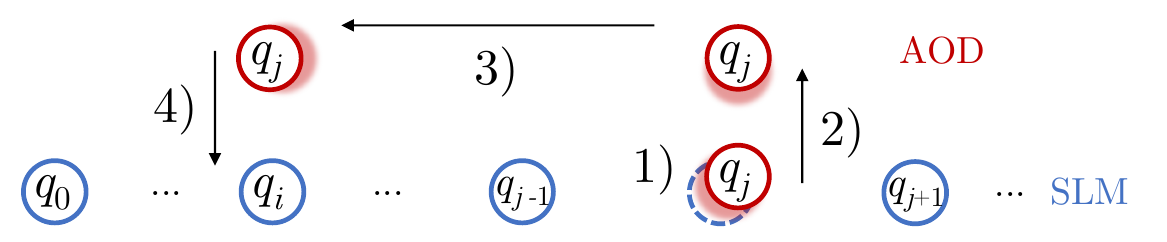}
    \caption{Universal QC with one AOD trap and one SLM row.
    Single-qubit gates executes directly.
    To implement an entangling gate on an arbitrary qubit pair: 1) pick $q_j$ from SLM to AOD, so the original SLM trap is empty (dashes), 2) move the AOD trap up, 3) adjust the AOD trap horizontally until $q_i$ and $q_j$ align, then 4) shift the AOD trap down to perform the gate.}
    \label{fig:universality}
\end{figure}

\subsection{Atom Transfer} \label{ssec:transfer}

So far, we have described atoms staying in their own individual tweezer traps, as was focused on in the experiments of \cite{nature22-lukim-bluvstein-atom-array}.
However, it has previously been demonstrated that atoms can be transferred between tweezer traps \cite{natphys07-beugnon-tweezer} by reducing the intensity of one tweezer trap while increasing or maintaining the intensity of another tweezer trap.
In the system considered here, in an AOD array, we can tune the individual intensity of AOD rows and columns to transfer to/from SLM traps: e.g., \autoref{fig:2}e, we turn off the leftmost AOD column so that $q_5$ is transferred to SLM.

\subsection{Universality} \label{ssec:universality}

With atom transfers, the architecture can perform universal quantum computing given a large enough area.
\autoref{fig:universality} depicts a toy construction.
We load the qubits to one SLM row with sufficient separations between the traps. 
There is one AOD trap working as delivery.
Per \autoref{impl:single-qubit}, single-qubit gates are always executable.
To apply an entangling gate on an arbitrary pair $(q_i, q_j)$, we perform the 4-step procedure illustrated in \autoref{fig:universality}.
Finally, we reverse the movements and put $q_j$ back to SLM.
Now, we are ready for the next gate.
With this construction, we can execute any single-qubit and two-qubit gate, so the architecture can perform universal QC.
Of course, this construction is like the demonstration of the Turing machine in classical universal computing where efficiency is not considered.
For example, we can easily put atoms in a square array that reduces the amount of time for movements.

\section{Quantum Layout Synthesis} \label{sec:mapping}

\begin{figure}[t]
    \centering
    \includegraphics[width=\linewidth]{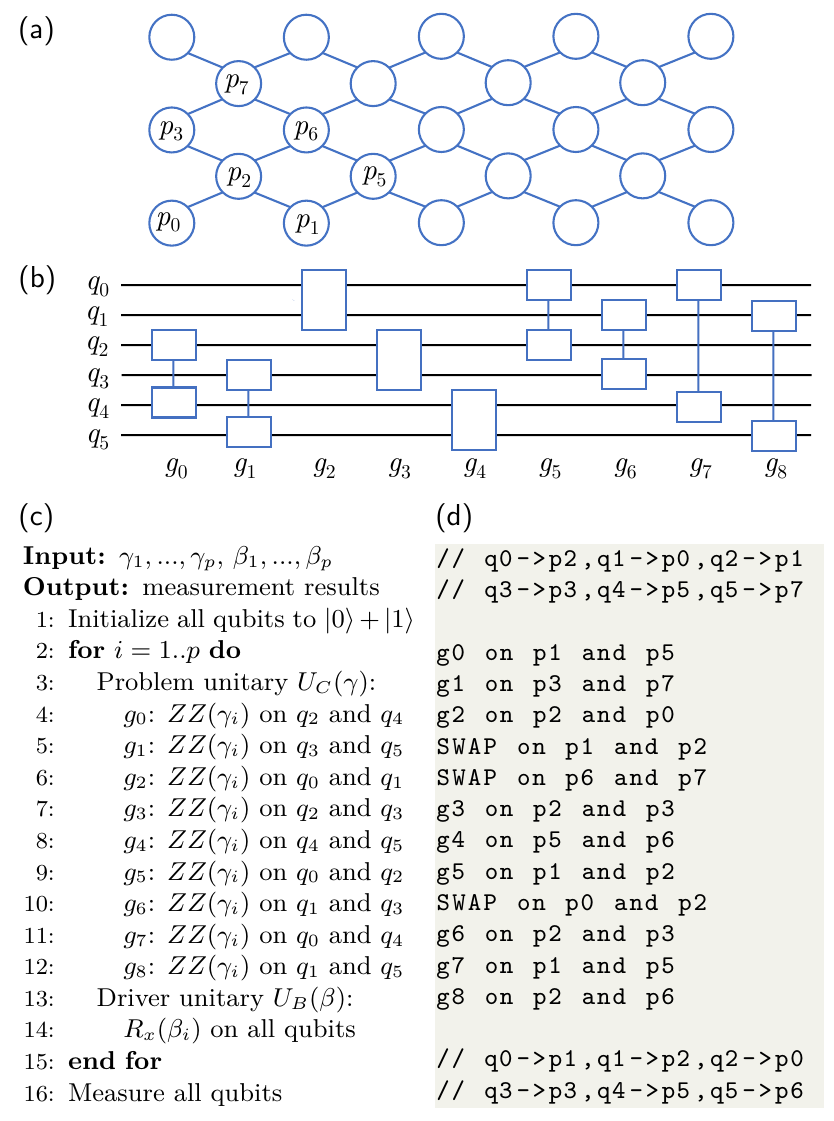}
    \caption{The quantum layout synthesis problem. 
    \textbf{(a)} The (partial) coupling graph of Google Sycamore processor, as in \cite{natphys21-google-qaoa}.
    The annotated quantum registers are made use of in this example.
    \textbf{(b)} Diagram of the quantum circuit to execute.
    \textbf{(c)} Pseudocode for QAOA applied to the Max-Cut problem.
    There are $p$ iterations of applying problem unitary $U_C$ and then driver unitary $U_B$.
    $U_C$ is implemented by the circuit in (b), though on different parameter $\gamma$ at different iterations.
    \textbf{(d)} A layout synthesis solution that runs circuit (b) on architecture (a). 
    The comments are the initial and final qubit mappings.}
    \label{fig:eg-problem}
\end{figure}

The input of quantum layout synthesis consists of two parts: the \textit{coupling graph} and the program/circuit to be executed.
For instance, the coupling graph of Google's Sycamore processor~\cite{web21-google-sycamore-data-sheet} is shown in \autoref{fig:eg-problem}a.
In this graph, every node is a \textit{quantum register} that can hold one qubit, and all possible two-qubit entangling operations/gates are represented by the edges.
An example of a quantum program is exhibited in \autoref{fig:eg-problem}c, which is the quantum approximate optimization algorithm (QAOA) \cite{arxiv1411-farhi-goldstone-gutmann-qaoa} applied to the Max-Cut problem of a 3-regular graph.
The parameters $\gamma$'s and $\beta$'s are inputs from an outer layer classical optimization.
The quantum program first initializes all qubits to state $|0\rangle+|1\rangle$ and iteratively applies the problem unitary $U_C(\gamma)$ and the driver unitary $U_B(\beta)$ for $p$ times, each iteration with different parameters $\gamma_i$ and $\beta_i$.
$U_B$ consists of single-qubit $R_x$ gates on all qubits which can always be executed, so it is not challenging for the compiler.
On the other hand, $U_C$, consisting of nine two-qubit ZZ-phase gates (\autoref{fig:eg-problem}b), is the center of attention in layout synthesis.
These gates are induced by the 3-regular graph: for each edge in the graph, we apply a ZZ-phase gate to its two qubits.

For this example, we first need to map the qubits $q_0$ ... $q_5$ to quantum registers, i.e., nodes in the coupling graph in \autoref{fig:eg-problem}a.
E.g., on the top of the solution shown in \autoref{fig:eg-problem}d, $q_2$ is initially mapped to $p_1$ and $q_4$ is mapped to $p_5$.
Since $p_1$ and $p_5$ are indeed adjacent in the coupling graph, the gate $g_0$ (on $q_2$ and $q_4$) can be executed.
However, $g_3$ is on $q_2$ and $q_3$ that are mapped to nonadjacent registers $p_1$ and $p_3$.
In this case, we need to insert a special gate named SWAP on $p_1$ and $p_2$ before $g_3$ that exchanges the two registers' qubits.
After the SWAP, $q_2$ maps to $p_2$ that is adjacent to $p_3$ which is holding $q_3$.
Then, $g_3$ can be executed under this updated mapping. 

Note that SWAP is not a native gate in most QC architectures.
To achieve its effect, we need three native entangling gates and a few single-qubit gates, whether the native entangling gate is a CNOT gate or a Cross-Resonance gate on IBM processors \cite{micro20-gokhale-javadi-abhari-earnest-shi-chong-compilation-openpulse}, or a $\sqrt{ i \mathrm{SWAP}}$ gate or a SYC gate on Sycamore \cite{natphys21-google-qaoa}.
Currently, the fidelity of entangling gates is low, so SWAPs are expensive.
Thus, layout synthesis \cite{dac23-olsq2, tc20-tan-cong-optimality-layout-queko, iccad20-tan-cong-optimal-layout-synthesis} is a very important, if not the most important, step in compilation as it determined the SWAPs to be inserted.
It is also referred to as qubit mapping \cite{ asplos19-li-ding-xie-sabre-mapping, aspdac19-zulehner-wille-su4-compiling, dac19-wille-burgholzer-zulehner-mapping-minimal-swaph,  iccad19-bhattacharjee-saki-alam-chattopadhyay-ghosh-muqut-mapping,  isca19-murali-linke-martonisi-abhari-nguyen-alderete-triq-architecture-studies, asplos21-zhang-hayes-qiu-jin-chen-zhang-time-optimal-mapping, iccad21-tan-cong-qubit-mapping-absorption}, qubit placement \cite{tcad08-maslov-falconer-mosca-placement, aspdac14-shafaei-saeedi-pedram-placement-communication-2d, arxiv1703-Bhattacharjee-Chattopadhyay-depth-optimal-placement}, or allocation \cite{cgo18-siraichi-santos-collange-pereira-qubit-allocation, dac19-ashsaki-alam-ghosh-qure-nisq, dac20-alam-ash-saki-ghosh-compilation-flow-qaoa}.
Minimizing the number of SWAPs, or the number of gate layers in layout synthesis is proven to be NP-hard \cite{tcad08-maslov-falconer-mosca-placement, tc20-tan-cong-optimality-layout-queko, cgo18-siraichi-santos-collange-pereira-qubit-allocation, socs18-botea-kishimoto-marinescu-complexity-quantum-compilation}.
Following the above approach, layout synthesis for DPQA is even more challenging because the coupling graph can change across different stages of the circuit execution, so we opt for a different formulation in this work.

DPQA is a relatively new technology, so previous works on compilation are for Rydberg atoms trapped in an SLM array that has fixed connectivity during the computation.
There are both experimental \cite{science22-lukin-mis} and theoretical/computational \cite{quantum22-pasqal-pulser, arxiv18-pichler-mis} works exploring solving the Max-Independent-Set problem using adiabatic quantum algorithm on SLM arrays.
Ref.~\cite{nature22-graham-atom-array} applies QAOA to the Max-Cut problem on an SLM array.
In this case, the layout synthesis problem is as if for a fixed architecture, like described above.
Ref.~\cite{isca22-patel-silver-tiwari-geyser-neutral} discusses logic synthesis for hypothetical architectures that can perform a three-qubit gate pulse sequence, but leverages existing layout synthesis tools for fixed architectures.
Ref.~\cite{isca21-baker-litteken-duckering-hoffmann-bernien-chong-long-distance-neutral} also utilizes existing layout synthesis tools but features the ability to perform long-range Rydberg interaction.
However, it requires a hypothetical architecture where the Rydberg range of qubits can be individually tuned. 

The most relevant previous work, Brandhofer et al.~\cite{iccad21-brandhoher-buchler-polian-optimal-mapping-atoms}, explores a reconfigurable but more constrained architecture.
In this specific architecture, qubits are arranged in a 2D grid with nearest-neighbor connectivity, and the assumption of local addressability for two-qubit gates is made.
The architecture permits `1D displacements,' allowing an entire row to shift left or right, altering the connectivity. However, this reconfiguration does not facilitate all-to-all connectivity, as qubits separated by multiple rows cannot be coupled through the 1D displacements.
In contrast, the 2D movements demonstrated in Ref.~\cite{nature22-lukim-bluvstein-atom-array} enable interactions between any two qubits in principle.
Importantly, the approach outlined in Ref.\cite{iccad21-brandhoher-buchler-polian-optimal-mapping-atoms} is intricately linked to its architecture assumptions, rendering it inapplicable in our case.

\section{Spatial Discretization of DPQA} \label{ssec:discrete-space}

\begin{figure}[t]
    \centering
    \includegraphics[width = \linewidth]{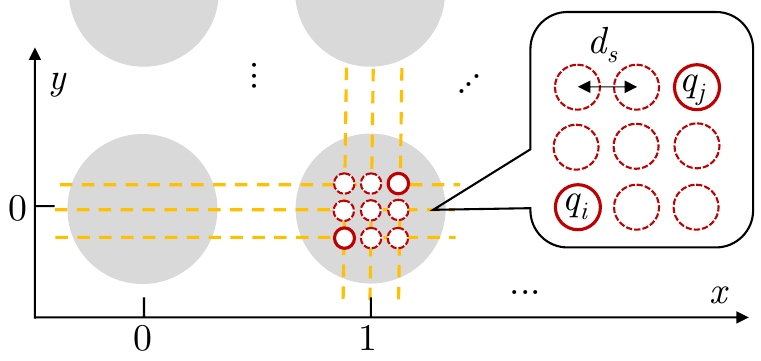}
    \caption{Discretization of space into interaction sites.
    The unit of X and Y is a sufficient distance to prevent Rydberg interaction.
    Interaction sites, indicated by shades, are centered at integer points on the 2D plane.
    A limited number of AOD rows or columns can stack together at one site.
    The callout is zooming into a site with three AOD rows and three columns.}
    \label{fig:discretization}
\end{figure}

There may be parallel executions of two-qubit entangling gates at different sites, so, per \autoref{impl:two-qubit-no-third}, the sites should be sufficiently separated to avoid unwanted Rydberg interactions.
Also, to maximize usage, the tiling pattern of the sites should accord to the geometry of the tweezer arrays, which is a 2D grid per \autoref{impl:optical-tweezer-array}.
The interaction sites are illustrated as shades in \autoref{fig:discretization}.
In fact, our efforts in discretization is analogous to that of Mead and Conway~\cite{book80-mead-conway-vlsi} in VLSI chip design where an abstract basic length unit in semiconductor fabrication, $\lambda$, was introduced.
The chip area is discretized to separated ``lines'' of $2\lambda$'s wide layout design.
These dimensionless $\lambda$-rules helped the advancement of automated layout tools despite the fast developments in the fabrication technology that affects $\lambda$.
Similarly, based on our discretization, our formulation holds even if the constants $r_b$, $2.5r_b$ and $d_s$ change.
It is crucial to retain this flexibility for possible adjustments in physics experiments.
E.g., we may want to excite the qubits to a higher Rydberg state, leading to a bigger $r_b$;
or upgrading to higher-resolution microscope objective lenses, leading to a smaller $d_s$.

We allow several rows or columns to ``stack'' together at one interaction site to support gates between two AOD qubits.
However, there is an upper bound on how many AOD rows/columns can be stacked together at a site because these AOD rows/columns cannot be too close to each other (\autoref{impl:row-separation}).
We denote the maximal stacking factor as $R_\mathrm{STK}$ and $C_\mathrm{STK}$, respectively.
They are decided by the minimal AOD row/column separation $d_s$ and the Rydberg range $r_b$.
The callout in \autoref{fig:discretization} exhibits an extreme case where we need to entangle two qubits $q_i$ and $q_j$ at the corners of the site.
This requires $[(R_\mathrm{STK}-1)^2+(C_\mathrm{STK}-1)^2]d_s^2\le r_b^2$.
$r_b=7.5\upmu$m and $d_s=2\upmu$m, $R_\mathrm{STK}=C_\mathrm{STK}=3$ satisfies the inequality.

In fact, the ticks on $x$ and $y$ axes in \autoref{fig:1}c and \autoref{fig:2}b-f indicates the interaction sites.
At each stage, per \autoref{impl:two-qubit-no-third}, there can be at most two qubits.
Thus, there are five possible situations at a site:
1) empty, e.g., (0,1) at stage 0 (\autoref{fig:2}b);
2) one SLM qubit, e.g., (1,1) at stage 2 holds only $q_4$;
3) one AOD qubit, e.g., (3,1) at stage 0 holds only $q_0$;
4) one SLM qubit and one AOD qubit, e.g, (1,1) at stage 0 holds $q_4$ and $q_2$;
5) two AOD qubits, e.g., (1,0) at stage 0 holds $q_5$ and $q_3$.

The discretized coordinates (of interactions sites) are enough to specify AOD and SLM qubit locations, but they are not sufficient as the state of the architecture because of the stacking of rows/columns we just mentioned.
For example, at stage 1 (\autoref{fig:2}c), both AOD rows are at $y=1$.
Because of \autoref{impl:aod-order}, the upper row cannot move across the lower row, e.g., $q_2$ cannot move below $q_3$.
With only coordinates, it is hard to enforce constraints like this.
Thus, as part of the architecture state, we also need to specify which row and column each AOD qubit is in.
Finally, we have to specify whether the qubit is in SLM or AOD at each stage to handle atom transfers. 

In conclusion, the computation progresses in multiple stages: stage 0, AOD movement 0, stage 1, AOD movement 1, stage 2, ...
At each stage, the architecture has a state consisting of interaction site indices (specifying location), AOD row/column indices, and an array index (specifying whether in SLM or AOD) for each qubit.
During the AOD movement, the AOD row/column indices and the array index are invariant, but the site indices can change as AOD traps move in space.

\section{SMT Constraints} \label{sec:formulation}
Constraints in this subsection come from physics implications on the architecture (circuit-independent), or fundamental properties of quantum programs (circuit-dependent).
Let us use $N$ for the number of qubits and $G$ for the number of gates.
Note that in the constraints below, we use `$=$' to denote the operation that returns Boolean true if the l.h.s equals  the r.h.s, and returns false otherwise.
`$[A,B)$' means from $A$ to $B-1$.
All the concrete examples are from \autoref{fig:2}, and the reader can plug in values from Appendix~\ref{ssec:json} for more examples.

\subsection{Circuit-Independent Constraints}
\textit{Upper bounding} the variables: $\forall i\in[0,N), s\in[0,S)$
\begin{equation}
    0\le x_{i,s}<X,
\end{equation}
similarly for $y$, $c$, and $r$ with bounds $Y$, $C$, and $R$.

\textit{Stationary SLM} enforces \autoref{impl:slm}: $\forall i\in[0,N),\ \forall s\in[0,S-1)$
\begin{equation}
    (a_{i,s}=0)\Rightarrow ((x_{i,s+1}=x_{i,s}) \wedge (y_{i,s+1}=y_{i,s})).
\end{equation}
E.g., $q_4$ is in SLM at stage 0, i.e., $a_{4,0}=0$, so its site indices remain the same between stage 0 and 1, i.e., $x_{4,1}=x_{4,0}$ and $y_{4,1}=y_{4,0}$.

\textit{AOD moves by whole rows/columns} enforcing \autoref{impl:aod-together}: $\forall i\in[0,N),\ \forall s\in[0,S-1)$
\begin{equation} \label{eq:aod-moves-by-whole}
    (a_{i,s}=1) \Rightarrow ((c_{i,s+1}=c_{i,s}) \wedge (r_{i,s+1}=r_{i,s})).
\end{equation}
E.g., $q_5$ is at column 0 at stage 0, $a_{5,0}=1$, so when it arrives at stage 1, the row/column index remains the same, i.e., $c_{5,1}=c_{5,0}$ and $r_{5,1}=r_{5,0}$, despite the changing site indices, i.e., $y_{5,1}\neq y_{5,0}$.

\textit{Site order implying row/column order} enforcing \autoref{impl:aod-order} in the case of non-stacking rows/columns: $\forall i,i'\in[0,N),\ \forall s\in[0,S)$
\begin{equation} \label{eq:site-order-row-order}
\begin{split}
    (x_{i,s}<x_{i',s})&\Rightarrow\ (c_{i,s}<c_{i',s}),\\
    (y_{i,s}<y_{i',s})&\Rightarrow (r_{i,s}<r_{i',s}).
\end{split}
\end{equation}
E.g., at stage 0, $q_5$ is at $x=1$ while $q_1$ is at $x=2$, so $c_{5,0}<c_{1,0}$.

\textit{No crossing between AOD row/columns} enforces \autoref{impl:aod-order} in the case of stacked rows/columns: $\forall i,i'\in[0,N),\ \forall s\in[0,S-1)$
\begin{equation}\begin{split}
    ((a_{i,s}=1) \wedge (a_{i',s}=1) \wedge &(c_{i,s}<c_{i',s}))\\
    &\Rightarrow (x_{i,s+1} \le x_{i', s+1}), \\
    ((a_{i,s}=1) \wedge (a_{i',s}=1) \wedge &(r_{i,s}<r_{i',s}))\\
    &\Rightarrow (y_{i,s+1} \le y_{i', s+1}). \\
    \end{split}
\end{equation}
E.g., at stage 0, $q_1$ is at row 0 and $q_0$ is at row 1, so $r_{1,0}<r_{0,0}$; at stage 1, $q_1$ and $q_0$ are both at $y=1$, which satisfies $y_{1,1}\le y_{0,1}$.

\textit{Maximal stacking} as in Appendix~\ref{ssec:discrete-space}: $\forall i,i'\in[0,N),\ \forall s\in[0,S)$
\begin{equation} \begin{split}
    ((a_{i,s-1}&=1) \wedge (a_{i',s-1}=1) \wedge  \\
    &(c_{i,s-1}-c_{i',s-1}\ge C_\mathrm{STK}))\Rightarrow (x_{i,s}>x_{i',s}), \\
    ((a_{i,s-1}&=1) \wedge (a_{i',s-1}=1) \wedge  \\
    &(r_{i,s-1}-r_{i',s-1}\ge R_\mathrm{STK}))\Rightarrow (y_{i,s}>y_{i',s}).
    \end{split}
\end{equation}
(When $s=0$, the $s-1$ above is 0; otherwise, it is $s-1$.)
E.g., at stage 0, $q_5$ is in column 0 while $q_0$ is in column 3, $c_{0,0}-c_{5,0}=3\ge C_\mathrm{STK}$, so $x_{0,1}>x_{5,1}$, i.e., they cannot be at the same site at stage 1.

\textit{One atom, one trap.}
There cannot be two atoms in one trap, thus imposing \autoref{impl:qubit-trap-mapping} and \ref{impl:two-qubit-no-third}.
If both atoms are in AOD, either their row or column index is different; if both are in SLM, either their site $x$ or $y$ index is different: $\forall i\in[0,N),\ \forall i'\in[i+1,N),\ \forall s\in[0,S)$
\begin{equation} \begin{split} \label{eq:one-atom-one-trap}
    ((a_{i,s}=1) &\wedge (a_{i',s}=1)\\
    &\Rightarrow ((c_{i,s}\neq c_{i',s})) \vee ((r_{i,s} \neq r_{i',s})), \\
    ((a_{i,s}=0) &\wedge (a_{i',s}=0)\\
    &\Rightarrow ((x_{i,s}\neq x_{i',s})) \vee ((y_{i,s} \neq y_{i',s})).
\end{split}
\end{equation}

\textit{(Optional) No atom transfer} by fixing array index (which is what we do in the evaluations for the optimal compiler): $\forall i\in[0,N),\ \forall s\in[0,S)$
\begin{equation}
    a_{i,s}=a_{i,0}.
\end{equation}
If it is allowed for an atom to transfer to an empty trap at a same site, i.e., forbidding transfer when there are two atoms at a site, $\forall i\in[0,N),\ \forall i'\in[i+1,N),\ \forall s\in[0,S-1)$
\begin{equation}
    \begin{split}
        ((x_{i,s+1}=&x_{i',s+1})\wedge(y_{i,s+1}=y_{i',s+1})) \\
        &\Rightarrow ((a_{i,s+1}=a_{i,s}) \wedge (a_{i',s+1}=a_{i',s})).
    \end{split}
\end{equation}

\subsection{Circuit-Dependent Constraints}
\textit{Gate collision.}
If two gates act on the same qubit, they cannot be executed at the same stage, e.g., $g_0$ and $g_3$ both act on $q_2$, so $t_0 \neq t_3$.

\textit{Gate dependence.} 
If the order of execution between two gates cannot be changed, we ensure this by $t_j<t_{j'}$ if $g_{j'}$ depends on $g_j$.

\textit{Connectivity} ensures \autoref{impl:two-qubit-close-necessary}.
Two qubits should be at the same site in order for an entangling gate to execute: $\forall j\in[0,G),$ $g_j$ acting on $q_i$ and $q_{i'}$, $\forall s\in [0,S)$
\begin{equation} \label{eq:connectivity}
    (t_j = s)\Rightarrow\ ((x_{i,s}=x_{i',s}) \wedge (y_{i,s}=y_{i',s})).
\end{equation}
E.g., $g_0$ at stage 0 is on $q_2$ and $q_4$, so $x_{2,0}=x_{4,0}$ and $y_{2,0}=y_{4,0}$.

\textit{Interaction exactness} enforces \autoref{impl:two-qubit-close-sufficient}.
We pre-compute a list $\rho_{i,i'}$ for each pair of qubits ($q_i$ and $q_{i'}$) that contains all the $j$ if $g_j$ acting on them.
In the example of \autoref{fig:2}, there is only one gate $g_2$ acting on $q_0$ and $q_1$, so $\rho_{0,1}=\{2\}$; in contrast, there is no gates on $q_0$ and $q_8$, so $\rho_{0,8}=\emptyset$.
If $\rho_{i,i'}\neq\emptyset$, then two qubits must be at the same site at some stage, and one of the gates on them is being executed at this stage: $\forall i\in[0,N),\ \forall i'\in[i+1,N),\ \mathrm{s.t. }\ \rho_{i,i'}\neq\emptyset,\ \forall s\in [0,S)$
\begin{equation} \label{eq:exactness}
    ((x_{i,s} = x_{i',s}) \wedge (y_{i,s} = y_{i',s})) \Rightarrow \left(\bigvee_{j\in \rho_{i,i'}} t_j=s\right).
\end{equation}
Conversely, if $\rho_{i,i'}=\emptyset$, the qubits should not be at the same site ever: $\forall i\in[0,N),\ \forall i'\in[i+1,N),\ \mathrm{s.t. }\ \rho(i,i')=\emptyset,\ \forall s\in [0,S)$
\begin{equation} \label{eq:exactness-empty}
    (x_{i,s} \neq x_{i',s}) \vee (y_{i,s} \neq y_{i',s}).
\end{equation}

\subsection{Enforcing Cardinality}
There are two ways to enforce cardinality: implicitly in variable definition or explicitly with a cardinality constraint.
The implicit approach is mainly for dimensions involved in the definition of the variables in the SMT model.
Our arrays of variables have two dimensions: the qubit and the stage, which means whatever the model can possibly express is a computation using that many qubits and that many stages.
The number of stages, $S$, in the optimal compiler is bounded in this approach: we only construct variables for $S$ stages.
If $S$ is too small to execute the whole circuit, the model is unsatisfiable, so we need to add more variables.
When the model becomes satisfiable, we have not introduced more variables than needed.
Considering the exponential scaling of SMT solving to model size, we opt for the implicit approach to force the cardinality of stages.

An example of the explicit approach appends the SMT model with a constraint like
\begin{equation}
   \sum_{j\in[0,G), s\in[S_\text{LB}, S_\text{UB})} \mathrm{ITE}(t_j=s,\ 1,\ 0) \ge M,
\end{equation}
where the stages between $S_\text{LB}$ and $S_\text{UB})$ are considered, and $\textrm{ITE}(\phi,w,z)$ means if the Boolean expression $\phi$ evaluates to true, return value $w$, otherwise $z$.
Essentially, the l.h.s.~is counting occurrences of a qubit pair appearing at the the same site at the same stage. 
If this sum is larger than $M$, then at least $M$ gates are executed between between stages $S_\text{LB}$ and $S_\text{LB}$.
There are many ways to decompose the above equation to Boolean logic.
We utilize the sequential counter approach offered by PySAT~\cite{imms-sat18} in the hybrid method (\autoref{fig:3}).
As a result, there are some intermediate Boolean variables introduced in the SMT model that do not correspond to any configurations of DPQA, purely for the sake of the cardinality constraint.

\subsection{Scalability of the Model}
The total number of variables in the optimal approach is $5NS+G$ where $N$ is the number of qubits, $S$ is the number of stages, and $G$ is the number of gates.
The total number of constraints is $O(G^2+GS+N^2S)$.
However, some of the variables have larger bounds.
If we represent the integer variables by bit-vectors, the total number of bits to represent the variables is $NS\log(2XYRC)+G\log(S)$, where $X$ and $Y$ are the dimensions of the interaction site grid, $C$ and $R$ are the number of AOD columns and rows.
The worst-case runtime of SMT solving is exponential, i.e., $O( (N_\mathrm{SLM} N_\mathrm{AOD})^{NS}\cdot S^G)$ where $N_\mathrm{SLM}=XY$ is the total number of SLM traps, and $N_\mathrm{AOD}$ is the total number of AOD traps.
In the shallow circuit regime where $S$ can be seen as a constant, and if the program is induced by sparse graphs so that $G=O(N)$, the number of bits required is $O(N\log(N_\mathrm{SLM} N_\mathrm{AOD}))$ and the number of constraints is $O(N^2)$.
For each `peeling' in the hybrid compiler, $S=2$.

\section{Evaluation Settings and Details} \label{sec:evaluation}
All the evaluation scripts are implemented in Python.
We used the following packages:
\texttt{pytket 1.13.2} which is the Python interface of the compiler t$|$ket$\rangle$~\cite{qst20-sivarajah-dikes-cowtan-simmons-edgington-duncan-tket-compiler-nisq};
and \texttt{qiskit 0.42.1} which is the Qiskit~\cite{Qiskit} release containing the compiler SABRE (originated from \cite{asplos19-li-ding-xie-sabre-mapping}).
Our compilers rely on a few Python packages.
The versions used during the evaluations are:
\texttt{z3-solver 4.12.1.0} which is the Z3 SMT solver (originated from \cite{tacas08-demoura-bjorner-z3-smt-solver});
\texttt{networkx 3.0} to calculate the maximum matching number of graphs;
\texttt{python-sat 0.1.8.dev1} to generate cardinality constraints;
and \texttt{matplotlib 3.6.2} to generate the figures.
The compilation appeared in the main text was ran on a desktop computer with an Intel Core i7-10700KF CPU and 32 GB RAM.
The compilation appeared in the appendices was ran on a server with two AMD EPYC 7V13 CPUs and 512 GB RAM.

In our compilers, we set the spatial bounds to $X=Y=R=C=16$.
For fairness of comparison, we assume the fixed architecture we are comparing with is equipped with the same gate set of DPQA.
The SWAP gate requires three two-qubit entangling gates and six single-qubit gates.
The benchmarks are graph circuits with 10, 12, 14, 16, 18, 20, 22, 30, 40, 50, 60, 70, 80, and 90 qubits.
We generated 10 3-regular graphs of each size.
For each graph, we assign a qubit to each node and append a two-qubit entangling gate for each pair of qubits connected by an edge to construct the graph circuit.
We set the time limit to $10^5$ seconds which is approximately a day.
Note that the compiler runtime can vary depending on the specific hardware and environment where it is run.
The timeout instances are $20_5$, $22_5$, and $22_8$ for the optimal approach, $80_1$, $90_0$, $90_2$, $90_6$, $90_8$, and $90_9$ for the hybrid approach, where the subscripts are the indices of the graph.
All the random graphs used are provided in the code base.

\section{Handling Generic Quantum Circuits}
\label{ssec:generic}

\begin{figure}[t]
    \centering
    \includegraphics[width=\linewidth]{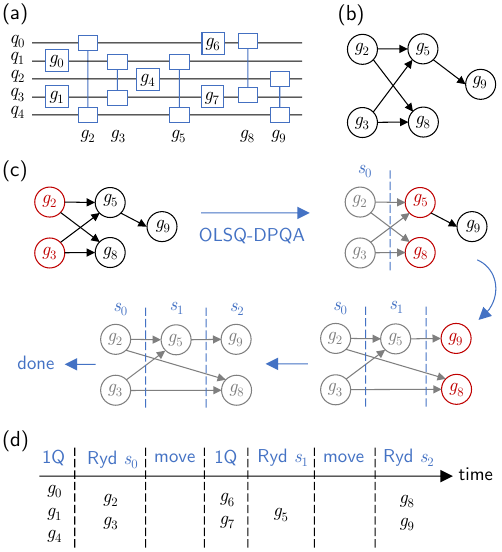}
    \caption{
    Handling generic circuits.
    \textbf{(a)} Example circuit.
    \textbf{(b)} Dependency graph of two-qubit gates.
    \textbf{(c)} Compilation process.
    OLSQ-DPQA is invoked 3 times.
    Each time, only the front layer (red nodes) is processed. 
    It is possible the entire front layer is not executed, leading to the inclusion of the remaining nodes in the subsequent front layer (e.g., $g_8$).
    \textbf{(d)} Final result.
    Prior to each Rydberg stage, we execute all single-qubit gates that has no dependency to any gates not yet executed.
    }
    \label{fig:generic}
\end{figure}

In the main text, our attention is primarily on the compilation of circuits comprised of commutable two-qubit gates.
We find that these circuits showcase the massive parallelism of DPQA architecture.
Also, the flexibility in commutation adds extra challenges to the compilation problems.
In \textit{generic circuits}, e.g., \autoref{fig:generic}a, there are two notable differences.
Firstly, these circuits include single-qubit gates (e.g., $g_0$ and $g_1$).
Secondly, the gates in these generic circuits are not necessarily commutable.
We assume a \textit{dependency} in cases where two gates act on the same qubit, dictating a fixed order; for instance, $g_0$ and $g_3$ both acting on $q_1$ means $g_3$ must be scheduled after $g_0$.
Our software implementation includes an \texttt{all\_commutable} flag as part of the problem specification.
When this flag is inactive, OLSQ-DPQA defaults to the workflow illustrated in \autoref{fig:generic}c: prior to compilation, we remove all single-qubit gates to derive the dependency graph of two-qubit gates, as shown in \autoref{fig:generic}b.
Due to the dependencies, only the front layer of the graph, represented by the red nodes (e.g., $g_2$ and $g_3$ initially), can be processed.
OLSQ-DPQA compiles the qubit movements for these gates, maximizing the number of executed gates, and removes them from the dependency graph (grayed out nodes).
Sometimes, not the entire front layer is executed depending on the qubit locations (e.g., $g_5$ is executed at $s_1$ while $g_8$ is not), leaving the remaining gates for the next round.
This process continues until all nodes are processed.
Finally, we reintroduce the single-qubit gates, as depicted in \autoref{fig:generic}d.
Prior to each two-qubit gate stage, we execute all single-qubit gates without dependencies at this point.
For instance, $g_7$ only depends on $g_3$, which is executed at $s_0$, allowing $g_7$ to be executed before $s_1$.

We benchmark OLSQ-DPQA on realistic generic circuits from QASMBench~\cite{li2022qasmbench}, detailed in \autoref{tab:benchmark}.
Specifically, we picked all the `medium' and `large' benchmarks with fewer than 100 qubits and less than 1000 gates.
Certain benchmarks share the same circuit family but differ in size, such as various-sized adders.
The \textit{2Q depth} of a circuit is the length of the longest path in the two-qubit dependency graph like \autoref{fig:generic}b.
For a fixed 10x10 grid qubit coupling graph, we utilized SABRE~\cite{asplos19-li-ding-xie-sabre-mapping} within Qiskit~\cite{Qiskit} to layout qubits and insert SWAPs.
In contrast, OLSQ-DPQA relies solely on qubit movement to route qubits, resulting in a reduction of two-qubit gates by 1.8X geomean, as shown in the rightmost column of \autoref{tab:benchmark}.
While, in most instances, the number of two-qubit stages (Rydberg) aligns closely with the 2Q depth of the circuit, OLSQ-DPQA may require a larger number of stages.
This arises from the fact that not all gates in the front layer can be executed at every stage due to the specific qubit locations at that point.
These front layers are generally less complicated than random graphs.
Consequently, even in cases where these benchmarks have more gates than graph circuits in the main text, the compiler runtime tends to be shorter.

\begin{table*}[p]
\caption{ \label{tab:benchmark} Compilation results of QASMBench~\cite{li2022qasmbench}.
We pick all of their `medium' and `large' benchmarks with less than 100 qubits and less than 1000 gates.
`2Q depth' of a circuit is the length of the longest path in the two-qubit dependency graph.
The number of Rydberg stages in the OLSQ-DPQA results is close to 2Q depth (1.13X geomean). 
The SABRE results assume a 10x10 grid qubit coupling graph.
OLSQ-DPQA reduces two-qubit gates because it uses movements instead of SWAPs to route qubits.}
\begin{center}
\begin{tabular}{|C{1.8cm} C{1cm} C{1.4cm} C{1.5cm}|C{1.6cm}|C{1.4cm} C{1.4cm} C{1.8cm}|C{1.8cm}|}
\hline
\multicolumn{4}{|c|}{\textbf{Benchmark statistics}} & \textbf{SABRE} & \multicolumn{3}{c|}{\textbf{OLSQ-DPQA}} & \textbf{Reduction} \\
Name   & Qubits & 2Q gates & 2Q depth & 2Q gates & 2Q gates & Rydbergs & Runtime/s & 2Q gates \\
\hline
\hline
seca       & 11   & 84    & 41        & 129       & 84            & 41       & 2.07E+1 & 1.54X          \\
sat        & 11   & 252   & 204       & 444       & 252           & 205      & 1.05E+2 & 1.76X          \\
cc         & 12   & 12    & 12        & 21        & 12            & 12       & 7.41E+0 & 1.75X          \\
multiply   & 13   & 40    & 23        & 64        & 40            & 23       & 1.63E+1 & 1.60X          \\
gcm          & 13 & 762 & 762 & 1257 & 762 & 762 & 5.27E+2 & 1.65X \\
bv         & 14   & 13    & 13        & 25        & 13            & 13       & 1.10E+1 & 1.92X          \\
qf21       & 15   & 115   & 112       & 202       & 115           & 112      & 1.07E+2 & 1.76X          \\
multiplier & 15   & 222   & 133       & 414       & 222           & 137      & 1.33E+2 & 1.86X          \\
dnn        & 16   & 384   & 48        & 384       & 384           & 53       & 6.47E+1 & 1.00X          \\
qec9xz     & 17   & 32    & 12        & 62        & 32            & 16       & 2.03E+1 & 1.94X          \\
qft        & 18   & 306   & 66        & 549       & 306           & 82       & 1.20E+2 & 1.79X          \\
bigadder   & 18   & 130   & 88        & 220       & 130           & 89       & 1.32E+2 & 1.69X          \\
square\_root & 18 & 898 & 644 & 1909 & 898 & 651 & 8.52E+2 & 2.13X \\
bv         & 19   & 18    & 18        & 39        & 18            & 18       & 2.88E+1 & 2.17X          \\
qram       & 20   & 136   & 80        & 247       & 136           & 82       & 1.49E+2 & 1.82X          \\
cat\_state & 22   & 21    & 21        & 39        & 21            & 21       & 4.76E+1 & 1.86X          \\
ghz\_state & 23   & 22    & 22        & 40        & 22            & 22       & 5.55E+1 & 1.82X          \\
swap\_test & 25   & 96    & 63        & 147       & 96            & 63       & 1.90E+2 & 1.53X          \\
knn        & 25   & 96    & 63        & 144       & 96            & 63       & 1.89E+2 & 1.50X          \\
ising      & 26   & 50    & 4         & 59        & 50            & 7        & 2.33E+1 & 1.18X          \\
wstate     & 27   & 52    & 28        & 67        & 52            & 28       & 1.02E+2 & 1.29X          \\
\hline
adder      & 28   & 195   & 97        & 321       & 195           & 98       & 3.90E+2 & 1.65X          \\
adder        & 64 & 455 & 181 & 845  & 455 & 188 & 8.32E+3 & 1.86X \\
bv         & 30   & 18    & 18        & 42        & 18            & 18       & 8.66E+1 & 2.33X          \\
bv         & 70   & 36    & 36        & 108       & 36            & 36       & 2.72E+3 & 3.00X          \\
cat        & 35   & 34    & 34        & 58        & 34            & 34       & 2.30E+2 & 1.71X          \\
cat        & 65   & 64    & 64        & 154       & 64            & 64       & 3.05E+3 & 2.41X          \\
cc         & 32   & 32    & 32        & 95        & 32            & 32       & 1.77E+2 & 2.97X          \\
cc         & 64   & 64    & 64        & 202       & 64            & 64       & 2.96E+3 & 3.16X          \\
dnn          & 33 & 248 & 95  & 365  & 248 & 97  & 5.59E+2 & 1.47X \\
dnn          & 51 & 392 & 140 & 632  & 392 & 152 & 2.91E+3 & 1.61X \\
ghz        & 40   & 39    & 39        & 87        & 39            & 39       & 3.86E+2 & 2.23X          \\
ghz        & 78   & 77    & 77        & 215       & 77            & 77       & 1.02E+4 & 2.79X          \\
ising      & 34   & 66    & 4         & 84        & 66            & 10       & 7.65E+1 & 1.27X          \\
ising      & 66   & 130   & 4         & 205       & 130           & 12       & 1.46E+3 & 1.58X          \\
ising        & 98 & 194 & 4   & 347  & 194 & 17  & 3.25E+4 & 1.79X \\
knn        & 31   & 120   & 78        & 186       & 120           & 78       & 4.07E+2 & 1.55X          \\
knn        & 67   & 264   & 168       & 486       & 264           & 168      & 8.14E+3 & 1.84X          \\
qft          & 29 & 812 & 110 & 1547 & 812 & 178 & 7.37E+2 & 1.91X \\
qugan      & 39   & 296   & 102       & 467       & 296           & 113      & 1.07E+3 & 1.58X          \\
qugan        & 71 & 552 & 182 & 936  & 552 & 193 & 1.36E+4 & 1.70X \\
swap\_test & 41   & 160   & 103       & 277       & 160           & 103      & 1.15E+3 & 1.73X          \\
swap\_test   & 83 & 328 & 208 & 628  & 328 & 208 & 3.71E+4 & 1.91X \\
wstate     & 36   & 70    & 37        & 94        & 70            & 37       & 3.01E+2 & 1.34X          \\
wstate     & 76   & 150   & 77        & 273       & 150           & 77       & 9.40E+3 & 1.82X      \\
\hline
\multicolumn{8}{|l|}{\textbf{\ geomean}} & 1.8X \\
\hline
\end{tabular}
\end{center}
\end{table*}

\section{SMT Values in the Running Example}
\label{ssec:json}

\begin{table}
\caption{\label{tab:t} Gate variable values in \autoref{fig:2}}
\begin{center}
\begin{tabular}{|C{1cm} |C{3cm} |C{0.6cm}|}
\hline
gate & qubits it acts on & $t$ \\
\hline
$g_0$ & $q_2$ and $q_4$ & 0 \\
\hline
$g_1$ & $q_3$ and $q_5$ & 0 \\
\hline
$g_2$ & $q_0$ and $q_1$ & 1 \\
\hline
$g_3$ & $q_2$ and $q_3$ & 1 \\
\hline
$g_4$ & $q_4$ and $q_5$ & 1 \\
\hline
$g_5$ & $q_0$ and $q_2$ & 2 \\
\hline
$g_6$ & $q_1$ and $q_3$ & 2 \\
\hline
$g_7$ & $q_0$ and $q_4$ & 3 \\
\hline
$g_8$ & $q_1$ and $q_5$ & 3 \\
\hline
\end{tabular}
\end{center}
\end{table}

\begin{table}
\caption{\label{tab:a} $a$ variable values in \autoref{fig:2}}
\begin{center}
\begin{tabular}{|C{0.6cm} |C{0.6cm} |C{0.6cm} |C{0.6cm}|C{0.6cm}|C{0.6cm}|C{0.6cm}|}
\hline
 & $q_0$ & $q_1$ & $q_2$ & $q_3$ & $q_4$ & $q_5$ \\
\hline
$s_0$ & 1 & 1 & 1 & 1 & 0 & 1 \\
\hline
$s_1$ & 1 & 1 & 1 & 1 & 0 & 1 \\
\hline
$s_2$ & 1 & 1 & 1 & 1 & 0 & 0 \\
\hline
$s_3$ & \textcolor{lightgray}{1} & \textcolor{lightgray}{1} & \textcolor{lightgray}{0} & \textcolor{lightgray}{1} & \textcolor{lightgray}{0} & \textcolor{lightgray}{1} \\
\hline
\end{tabular}
\end{center}
\end{table}

\begin{table}
\caption{\label{tab:x} $x$ variable values in \autoref{fig:2}}
\begin{center}
\begin{tabular}{|C{0.6cm} |C{0.6cm} |C{0.6cm} |C{0.6cm}|C{0.6cm}|C{0.6cm}|C{0.6cm}|}
\hline
 & $q_0$ & $q_1$ & $q_2$ & $q_3$ & $q_4$ & $q_5$ \\
\hline
$s_0$ & 3 & 2 & 1 & 1 & 1 & 1 \\
\hline
$s_1$ & 3 & 3 & 2 & 2 & 1 & 1 \\
\hline
$s_2$ & 2 & 2 & 2 & 2 & 1 & 1 \\
\hline
$s_3$ & 1 & 1 & 0 & 0 & 1 & 1 \\
\hline
\end{tabular}
\end{center}
\end{table}

\begin{table}
\caption{\label{tab:y} $y$ variable values in \autoref{fig:2}}
\begin{center}
\begin{tabular}{|C{0.6cm} |C{0.6cm} |C{0.6cm} |C{0.6cm}|C{0.6cm}|C{0.6cm}|C{0.6cm}|}
\hline
 & $q_0$ & $q_1$ & $q_2$ & $q_3$ & $q_4$ & $q_5$ \\
\hline
$s_0$ & 1 & 0 & 1 & 0 & 1 & 0 \\
\hline
$s_1$ & 1 & 1 & 1 & 1 & 1 & 1 \\
\hline
$s_2$ & 1 & 0 & 1 & 0 & 1 & 0 \\
\hline
$s_3$ & 1 & 0 & 1 & 0 & 1 & 0 \\
\hline
\end{tabular}
\end{center}
\end{table}

\begin{table}
\caption{\label{tab:c} $c$ variable values in \autoref{fig:2}}
\begin{center}
\begin{tabular}{|C{0.6cm} |C{0.6cm} |C{0.6cm} |C{0.6cm}|C{0.6cm}|C{0.6cm}|C{0.6cm}|}
\hline
 & $q_0$ & $q_1$ & $q_2$ & $q_3$ & $q_4$ & $q_5$ \\
\hline
$s_0$ & 3 & 2 & 1 & 1 & \textcolor{lightgray}{2} & 0 \\
\hline
$s_1$ & 3 & 2 & 1 & 1 & \textcolor{lightgray}{1} & 0 \\
\hline
$s_2$ & 3 & 2 & 1 & 1 & \textcolor{lightgray}{1} & \textcolor{lightgray}{1} \\
\hline
$s_3$ & \textcolor{lightgray}{3} & \textcolor{lightgray}{2} & \textcolor{lightgray}{1} & \textcolor{lightgray}{1} & \textcolor{lightgray}{2} & \textcolor{lightgray}{3} \\
\hline
\end{tabular}
\end{center}
\end{table}

\begin{table}
\caption{\label{tab:r} $r$ variable values in \autoref{fig:2}}
\begin{center}
\begin{tabular}{|C{0.6cm} |C{0.6cm} |C{0.6cm} |C{0.6cm}|C{0.6cm}|C{0.6cm}|C{0.6cm}|}
\hline
 & $q_0$ & $q_1$ & $q_2$ & $q_3$ & $q_4$ & $q_5$ \\
\hline
$s_0$ & 1 & 0 & 1 & 0 & \textcolor{lightgray}{1} & 0 \\
\hline
$s_1$ & 1 & 0 & 1 & 0 & \textcolor{lightgray}{0} & 0 \\
\hline
$s_2$ & 1 & 0 & 1 & 0 & \textcolor{lightgray}{0} & \textcolor{lightgray}{1} \\
\hline
$s_3$ & \textcolor{lightgray}{1} & \textcolor{lightgray}{0} & \textcolor{lightgray}{1} & \textcolor{lightgray}{0} & \textcolor{lightgray}{1} & \textcolor{lightgray}{1} \\
\hline
\end{tabular}
\end{center}
\end{table}

We provide the values of all the SMT variables in the running example illustrated in \autoref{fig:2}.
This example includes 9 gates on 6 qubits.
\autoref{tab:t} provides the qubits each gate acts on along with the time coordinates of these gates.
\autoref{tab:a} provides the array index of each qubit at each stage.
We make the values for the last stage gray because they do not affect the solution in any way.
(Note that for $S$ stages, there are only $S-1$ movements in between; and our convention is that the movement between $s_i$ and $s_{i+1}$ is encoded in the $a$, $c$, and $r$ variables of $s_i$.)
\autoref{tab:x} and \autoref{tab:y} provides the site indices.
In this example, $X=4$ and $Y=2$, meaning all $x$ variables are in $\{0,1,2,3\}$ and all $y$ variables are in $\{0,1\}$.
\autoref{tab:c} and \autoref{tab:r} provides the AOD column and row indices.
In this example, $C=4$ and $R=2$.
Apart from the values for the last stage, some other values are also gray because the corresponding qubit is in SLM in that stage.
For convenience in comparing with \autoref{fig:2}, we also reorganize the values based on stages as follows.
\begin{lstlisting}[style=base]
stage0: [
  {qubit: 0, a: 1, x: 3, y: 1, c: 3, r: 1},
  {qubit: 1, a: 1, x: 2, y: 0, c: 2, r: 0},
  {qubit: 2, a: 1, x: 1, y: 1, c: 1, r: 1},
  {qubit: 3, a: 1, x: 1, y: 0, c: 1, r: 0},
  {qubit: 4, a: 0, x: 1, y: 1, @c: 2@, @r: 1@},
  {qubit: 5, a: 1, x: 1, y: 0, c: 0, r: 0} ];
stage1: [
  {qubit: 0, a: 1, x: 3, y: 1, c: 3, r: 1},
  {qubit: 1, a: 1, x: 3, y: 1, c: 2, r: 0},
  {qubit: 2, a: 1, x: 2, y: 1, c: 1, r: 1},
  {qubit: 3, a: 1, x: 2, y: 1, c: 1, r: 0},
  {qubit: 4, a: 0, x: 1, y: 1, @c: 1@, @r: 0@},
  {qubit: 5, a: 1, x: 1, y: 1, c: 0, r: 0} ];
stage2: [
  {qubit: 0, a: 1, x: 2, y: 1, c: 3, r: 1},
  {qubit: 1, a: 1, x: 2, y: 0, c: 2, r: 0},
  {qubit: 2, a: 1, x: 2, y: 1, c: 1, r: 1},
  {qubit: 3, a: 1, x: 2, y: 0, c: 1, r: 0},
  {qubit: 4, a: 0, x: 1, y: 1, @c: 1@, @r: 0@},
  {qubit: 5, a: 0, x: 1, y: 0, @c: 1@, @r: 1@} ];
stage3: [
  {qubit: 0, @a: 1@, x: 1, y: 1, @c: 3@, @r: 1@},
  {qubit: 1, @a: 1@, x: 1, y: 0, @c: 2@, @r: 0@},
  {qubit: 2, @a: 0@, x: 0, y: 1, @c: 1@, @r: 1@},
  {qubit: 3, @a: 1@, x: 0, y: 0, @c: 1@, @r: 0@},
  {qubit: 4, @a: 0@, x: 1, y: 1, @c: 2@, @r: 1@},
  {qubit: 5, @a: 1@, x: 1, y: 0, @c: 3@, @r: 1@} ].
\end{lstlisting}

\end{document}